\documentclass[manuscript,screen]{acmart}

\AtBeginDocument{%
  \providecommand\BibTeX{{%
    \normalfont B\kern-0.5em{\scshape i\kern-0.25em b}\kern-0.8em\TeX}}}

\setcopyright{acmcopyright}
\copyrightyear{TBD}
\acmYear{TBD}
\acmDOI{TBD}

\usepackage{subcaption}
\usepackage[ruled,vlined,noend,linesnumbered]{algorithm2e}
\usepackage{calrsfs}
\usepackage{pgfplots}
\usepackage{mathtools}
\usepackage{multirow}
\usepackage{fancyvrb}
\usepackage{setspace}
\usepackage[capitalize]{cleveref}

\DeclareMathAlphabet{\pazocal}{OMS}{zplm}{m}{n}
\newcommand{\ten}[1]{\pazocal{#1}}

\newcommand{\norm}[1]{\left\lVert#1\right\rVert}
\newcommand\inputpgf[2]{{
\let\pgfimageWithoutPath\pgfimage
\renewcommand{\pgfimage}[2][]{\pgfimageWithoutPath[##1]{#1/##2}}
\let\includegraphicsWithoutPath\includegraphics
\renewcommand{\includegraphics}[2][]{\includegraphicsWithoutPath[##1]{#1/##2}}
\input{#1/#2}
}}
\SetArgSty{textnormal}

\ifdefined\modlevelthreshold \else \def\modlevelthreshold{100} \fi
\newcounter{modlevel}
\newcommand{\modified}[2][2]{\setcounter{modlevel}{#1}\ifnum\value{modlevel}>\modlevelthreshold{\color{blue}#2}\else#2\fi}
\newcommand{\mtprod}[1]{\cdot_{#1}}
\newcommand{\tuckerprod}{\cdot}

\newcommand{\tomsonly}[2]{\ifdefined\toms#1\else#2\fi}
\newcommand{\pathsuffix}{\tomsonly{/acm-toms}{/acm-toms}}

\definecolor{dark-green}{rgb}{0, 0.5, 0}

\allowdisplaybreaks
\begin{document}

\title{\tomsonly{Algorithm xxxx: ATC,}{ATC:} an Advanced Tucker Compression library for multidimensional data}

\author{Wouter Baert}
\email{wouter.baert@kuleuven.be}
\orcid{0000-0002-2846-9199}
\affiliation{
  \institution{KU Leuven}
  \department{Department of Computer Science}
  \streetaddress{Celestijnenlaan 200A -- bus 2402}
  \city{Leuven}
  \postcode{3001}
  \country{Belgium}
}

\author{Nick Vannieuwenhoven}
\email{nick.vannieuwenhoven@kuleuven.be}
\orcid{0000-0001-5692-4163}
\affiliation{
  \institution{KU Leuven}
  \department{Department of Computer Science}
  \streetaddress{Celestijnenlaan 200A -- bus 2402}
  \city{Leuven}
  \postcode{3001}
  \country{Belgium}
}

\begin{abstract}
We present ATC, a C++ library for advanced Tucker-based \modified[1]{lossy} compression of \modified[1]{dense} multidimensional numerical data \modified[1]{in a shared-memory parallel setting}, based on the sequentially truncated higher-order singular value decomposition (ST-HOSVD) and bit plane truncation. Several techniques are proposed to improve \modified{speed, memory usage, error control and compression rate.} First, a hybrid truncation scheme is described which combines Tucker rank truncation and TTHRESH quantization [Ballester-Ripoll et al., IEEE Trans. Visual. Comput. Graph., 2020]. We derive a novel expression to approximate the error of truncated Tucker decompositions in the case of core and factor perturbations. Furthermore, \modified{we parallelize the quantization and encoding scheme and adjust this phase to improve error control.} Moreover, implementation aspects are described, such as an ST-HOSVD procedure using only a single transposition. We also discuss several usability features of ATC, including the presence of multiple interfaces, extensive data type support and integrated downsampling of the decompressed data. Numerical results show that ATC maintains state-of-the-art Tucker compression rates, while providing average speed-up factors of 2.2-\modified[4]{3.5} and halving memory usage. Furthermore, our compressor provides precise error control, only deviating 1.4\% from the requested error on average. \modified[1]{Finally, ATC often achieves higher compression than non-Tucker-based compressors in the high-error domain.}
\end{abstract}



\maketitle

\section{Introduction}

Many scientific, industrial and medical applications generate numerical data on multidimensional grids, such as hyperspectral imaging \cite{moffett-field}, diffusion tensor imaging \cite{brain}, X-ray scans \cite{foot} and simulations of various kinds \cite{jhtdb,deforest-globe,hurricane-isabel}. As these datasets increase in size, so does the need for compression to reduce network and storage costs. Specifically, when dealing with floating-point data, lossy compression is most applicable because there is often no need to store the data in full precision. In fact, the original data may already be subject to measurement, simulation and round-off errors of a certain magnitude, making storage beyond this level of accuracy redundant.

\begin{figure}[t]
    \centering
    \begin{subfigure}[t]{0.25\textwidth}
        \centering
        \includegraphics[width=0.9\textwidth]{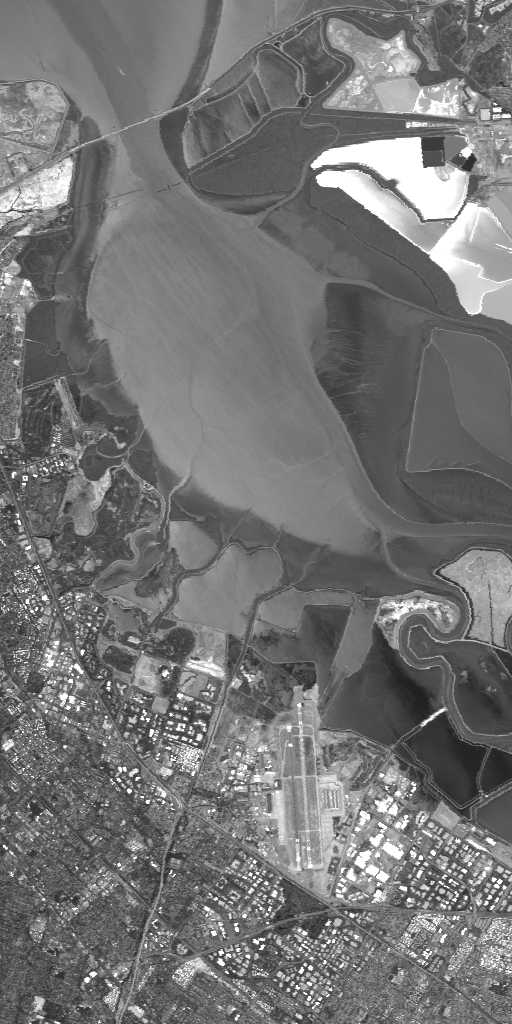}
        \caption{Slice 28}
    \end{subfigure}
    ~
    \begin{subfigure}[t]{0.25\textwidth}
        \centering
        \includegraphics[width=0.9\textwidth]{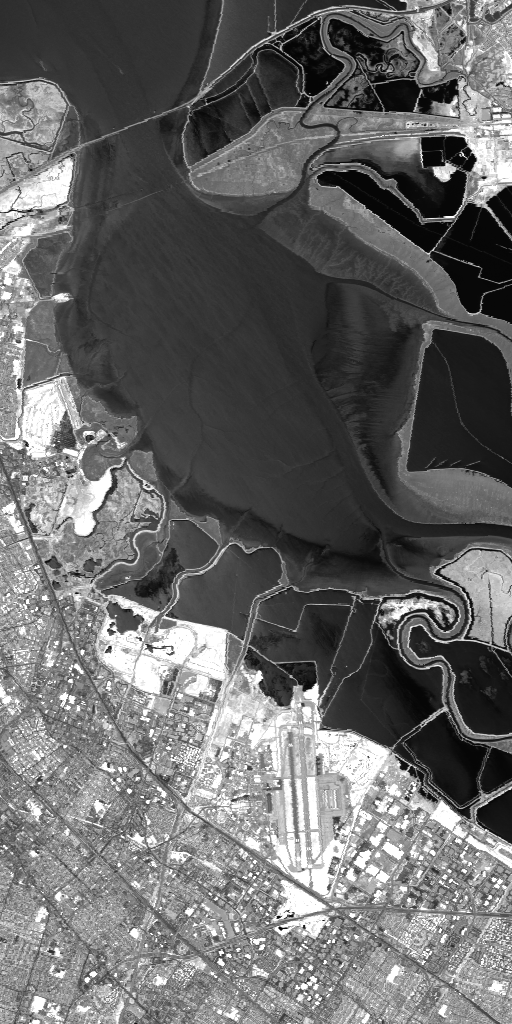}
        \caption{Slice 84}
    \end{subfigure}
    ~
    \begin{subfigure}[t]{0.25\textwidth}
        \centering
        \includegraphics[width=0.9\textwidth]{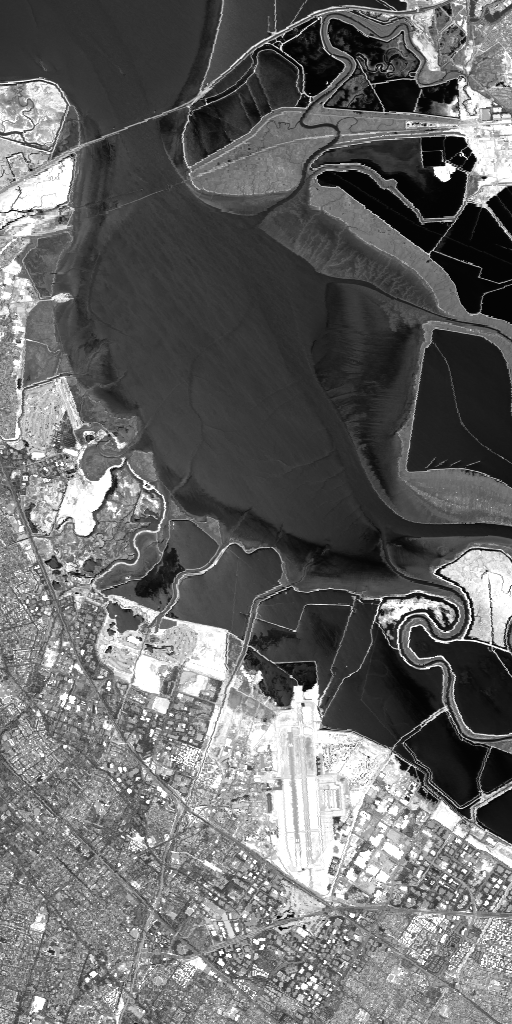}
        \caption{Slice 140}
    \end{subfigure}
    ~
    \begin{subfigure}[t]{0.25\textwidth}
        \centering
        \includegraphics[width=0.9\textwidth]{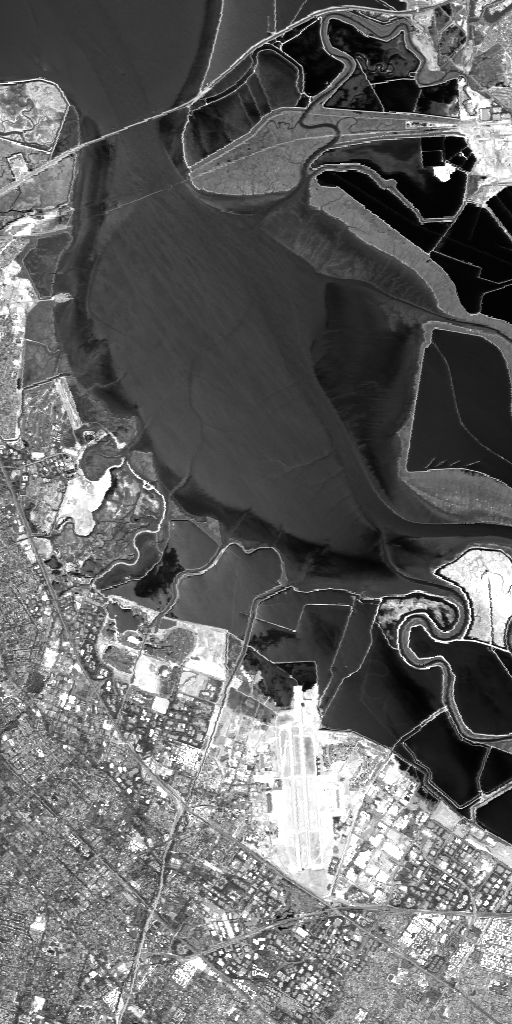}
        \caption{Slice 196}
    \end{subfigure}
    \caption{A sample of spectral slices from the Moffett-Field hyperspectral image (indices are 0-based). For visibility, a different scaling factor was used for each visualization, yet all slices clearly exhibit the same structure, showing great potential for compression.}
    \label{fig:data-slices-example}
\end{figure}

\begin{figure}[t]
    \centering
    \begin{subfigure}[t]{0.25\textwidth}
        \centering
        \includegraphics[width=0.9\textwidth]{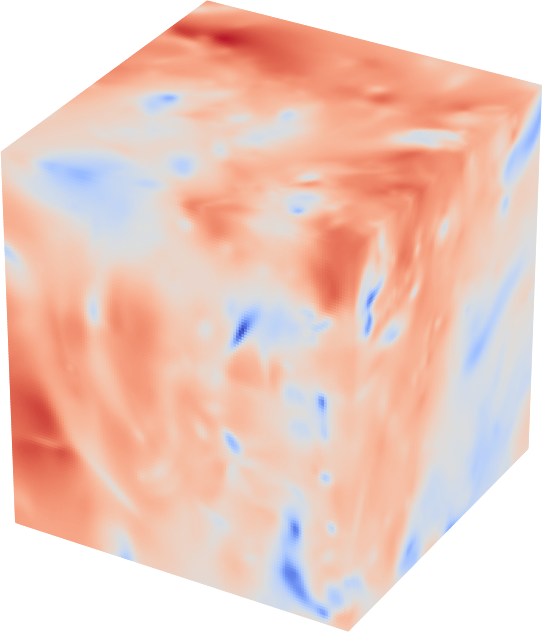}
    	\captionsetup{justification=centering,labelformat=empty}
        \caption{\modified{
        	Original
        }}
    \end{subfigure}
    ~
    \begin{subfigure}[t]{0.25\textwidth}
        \centering
        \includegraphics[width=0.9\textwidth]{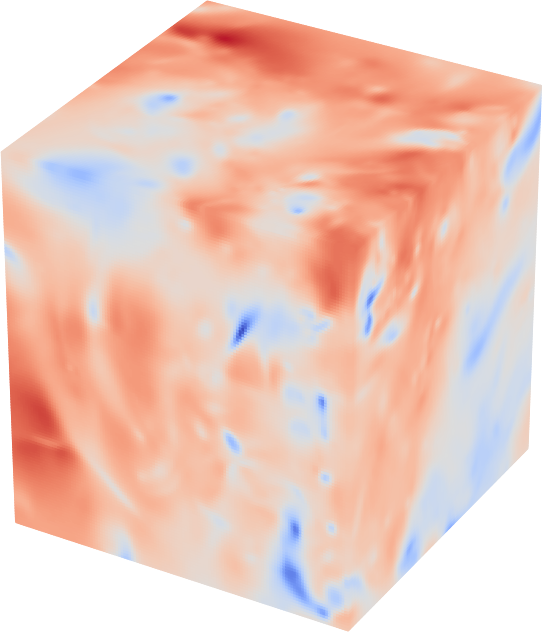}
    	\captionsetup{justification=centering,labelformat=empty}
        \caption{\modified{
        	Relative error: 0.100\%
        	\\ Compression factor: \modified[4]{389.28}
        }}
    \end{subfigure}
    ~
    \begin{subfigure}[t]{0.25\textwidth}
        \centering
        \includegraphics[width=0.9\textwidth]{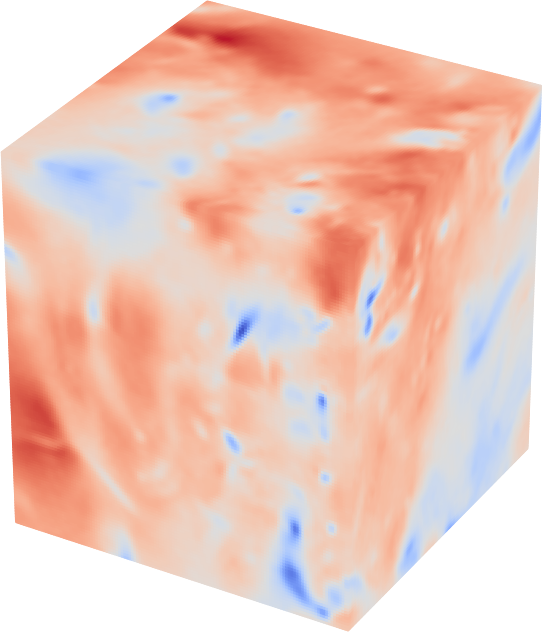}
    	\captionsetup{justification=centering,labelformat=empty}
        \caption{\modified{
        	Relative error: 1.00\%
        	\\ Compression factor: \modified[4]{3568.6}
        }}
    \end{subfigure}
    ~
    \begin{subfigure}[t]{0.25\textwidth}
        \centering
        \includegraphics[width=0.9\textwidth]{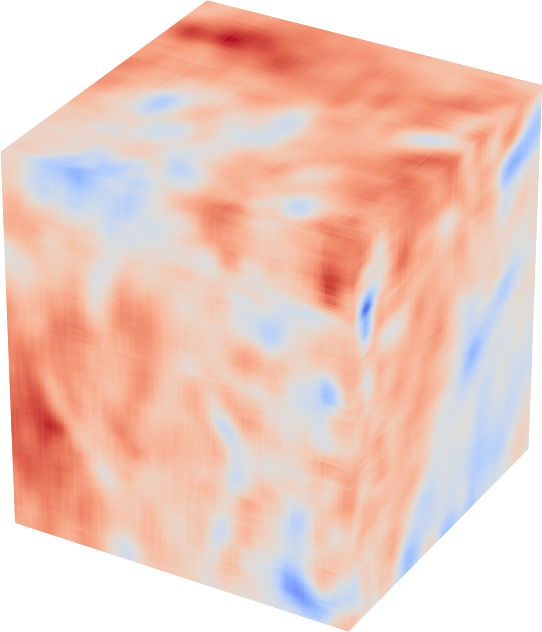}
    	\captionsetup{justification=centering,labelformat=empty}
        \caption{\modified[4]{
        	Relative error: 10.5\%
        	\\ Compression factor: 95034
        }}
    \end{subfigure}
    \caption{ATC compression examples using the Isotropic-PT dataset. Each visualization only shows the first time slice of the data tensor, while the statistics in the captions represent the full data.}
    \label{fig:compression-example}
\end{figure}

However, simply storing each data value within this limited precision is usually not the best approach. As shown in \cref{fig:data-slices-example}, realistic datasets often contain a lot of redundancy which can be removed during compression. In particular, when handling smooth datasets such as the one presented in \cref{fig:compression-example}, high compression factors can often be achieved at a small cost in compression error.

In this paper we propose ATC \modified[1]{(the Advanced Tucker Compressor)}, a lossy \modified[1]{compression library} for multidimensional numerical data, based on the \modified[1]{sequentially truncated higher-order singular value decomposition} (ST-HOSVD) \cite{st_hosvd} and an existing bit-plane-truncation-based quantization and encoding scheme from the TTHRESH compressor \cite{tthresh}. Our motivation is twofold: first, we aim to improve \modified{speed, memory usage, error control and compression rate} where possible compared to existing Tucker-based \cite{tucker} methods \modified{such as TTHRESH \modified[1]{\cite{tthresh}} and TuckerMPI \cite{tuckermpi}}. \modified{Specifically, we are interested in the compression of general, dense tensors, such as grid-based \modified[1]{physical measurements} or simulation data.} Second, our objective is to create a \modified{Tucker-based compression library with state-of-the-art performance and usability features such as multiple interfaces, broad support for data types\modified[1]{, shared-memory parallelism} and both file-based and in-memory compression or decompression.}

\modified{\modified[1]{In our view, these broad features are not yet available in existing \modified[4]{tensor-based compressors}. For example,} TTHRESH lacks most of these features as well as a programming interface and can therefore be used only as an executable instead of a library. On the other hand, \modified[1]{while TuckerMPI supports distributed-memory parallelism, it} is solely aimed at computing Tucker decompositions and as such achieves much worse compression rates, due to a lack of quantization and encoding methods; see \cref{tab:incremental-performance}.}

\begin{table}[]
\centering\small
\begin{tabular}{l|rrrrrrrrr}
\multicolumn{1}{c|}{\multirow{2}{*}{Compressor}} & \multicolumn{1}{c}{\multirow{2}{*}{ \begin{tabular}{@{}c@{}} Rel. \\ comp. \end{tabular} }} & \multicolumn{2}{c}{Speed-up}                            & \multicolumn{4}{c}{Peak memory (bytes/element)}                   & \multicolumn{2}{c}{\multirow{2}{*}{ \begin{tabular}{@{}c@{}} Error \\ control \end{tabular} }} \\
\multicolumn{1}{c|}{}                            & \multicolumn{1}{c}{}                                   & \multicolumn{1}{c}{Comp.} & \multicolumn{1}{c}{Decomp.} & \multicolumn{2}{c}{Comp.} & \multicolumn{2}{c}{Decomp.} & \multicolumn{2}{c}{}                               \\ \hline
\textbf{TTHRESH} & - & - & - & 40.3 & - & 38.7 & - & 33.8\% & - \\ \hline
\textbf{Baseline ATC} & +0.8\% & \textcolor{dark-green}{+27\%} & -0.8\% & 19.1 & \textbf{\textcolor{dark-green}{-53\%}} & 16.5 & \textbf{\textcolor{dark-green}{-57\%}} & 34.4\% & \textcolor{red}{+1.7\%} \\
+ Rank truncation & \textcolor{red}{-8.2\%} & \textbf{\textcolor{dark-green}{+54\%}} & \textcolor{dark-green}{+33\%} & 18.1 & \textcolor{dark-green}{-5.0\%} & 16.5 & +0.0\% & 25.8\% & \textbf{\textcolor{dark-green}{-25\%}} \\
+ Parallel quant./encoding  & -0.4\% & \textbf{\textcolor{dark-green}{+77\%}} & \textbf{\textcolor{dark-green}{+61\%}} & 18.4 & \textcolor{red}{+1.5\%} & 16.8 & \textcolor{red}{+1.4\%} & 27.7\% & \textcolor{red}{+7.0\%} \\
+ Predict dequant. correction & \textcolor{dark-green}{+4.0\%} & -3.1\% & +1.5\% & 18.4 & -0.2\% & 16.8 & -0.1\% & 1.4\% & \textbf{\textcolor{dark-green}{-95\%}} \\
+ Householder compression & \textcolor{dark-green}{+1.2\%} & +5.7\% & +1.9\% & 18.6 & \textcolor{red}{+1.2\%} & 16.7 & -0.1\% & 1.4\% & \textcolor{dark-green}{-1.1\%} \\
+ Mode storage order heuristic & \textcolor{dark-green}{+1.1\%} & -0.2\% & +2.5\% & 18.6 & -0.1\% & 16.9 & +0.9\% & 1.9\% & \textbf{\textcolor{red}{+35\%}} \\
+ Split bit plane truncation & \textcolor{dark-green}{+3.0\%} & -1.4\% & -1.8\% & 18.6 & -0.1\% & 16.9 & +0.0\% & 1.4\% & \textbf{\textcolor{dark-green}{-23\%}} \\
\textbf{Full ATC} & \textcolor{dark-green}{+1.1\%} & \textbf{\textcolor{dark-green}{+249\%}} & \textbf{\textcolor{dark-green}{+121\%}} & 18.6 & \textbf{\textcolor{dark-green}{-54\%}} & 16.9 & \textbf{\textcolor{dark-green}{-56\%}} & 1.4\% & \textbf{\textcolor{dark-green}{-96\%}} \\ \hline
\textbf{TuckerMPI} & \textbf{\textcolor{red}{-90\%}} & \textbf{\textcolor{dark-green}{+308\%}} & \textbf{\textcolor{dark-green}{+135\%}} & 10.8 & \textbf{\textcolor{dark-green}{-73\%}} & 9.6 & \textbf{\textcolor{dark-green}{-75\%}} & 86.2\% & \textbf{\textcolor{red}{+155\%}}
\end{tabular}

\vspace{5pt}
\caption{\modified[4]{Relative compression factors, compression and decompression times, peak memory usage per data element as well as deviations from the target error, averaged out over a range of compression errors and datasets. Error control is expressed as the average relative deviation of the actual compression error from the target error. Each boldface compressor shows relative performance compared to TTHRESH, while the other lines show incremental performance gains (green) or losses (red) achieved by a certain compressor optimization compared to the last line. For memory usage and error control, we also show the absolute performance metrics. Note that these metrics are averaged over a wide variety of settings, with certain optimizations having very different effects depending on the situation. The full methodology used is described in \cref{sec:results}.}}
\label{tab:incremental-performance}
\end{table}

\modified{\Cref{tab:incremental-performance} summarizes the performance gains achieved by the ATC optimizations compared to the other Tucker-based compressors TTHRESH and TuckerMPI. Thanks to an efficient implementation, e.g., avoiding unnecessary copies, baseline ATC already achieves a roughly $\modified[4]{27\%}$ compression speed-up and \modified[4]{roughly halves} memory usage reduction compared to TTHRESH, while being quite similar. Furthermore, the various algorithmic improvements made by ATC accumulate into an average compression and decompression speed-up factor of $\modified[4]{3.5}$ and $2.2$ respectively as well as $24$ times more precise error control, while maintaining similar compression rates. Moreover, the user can make certain trade-offs, for example increasing compression while lowering speed by reducing the level of rank truncation. In summary, this shows the significant improvements made by ATC over existing work, both algorithmically and in terms of implementation.}

In this paper, we will discuss the following aspects. First, we describe several general compression techniques and alternative compressors (\cref{sec:related-work}) as well as preliminary material that will be used throughout the paper (\cref{sec:preliminaries}). Then, we elaborate on our main algorithmic contributions (\cref{sec:algorithmic_improvements}). Afterwards, we summarize specific implementation aspects of the software (\cref{sec:implementation}). Finally, numerical results will be presented and discussed (\cref{sec:results}), \modified[1]{followed by concluding remarks and potential future lines of research (\cref{sec:conclusions}). In the supplementary materials, we discuss additional minor improvements and features.}

\section{Related work}
\label{sec:related-work}

In the past decades, a variety of methods were invented to address compression needs in various application domains \cite{data-compression}. In this section \modified[1]{we discuss the most relevant related methods.}

\subsection{\modified{Compression basics}}
\label{sec:compression-basics}

\modified{In data compression, methods are often divided into two categories: \emph{lossless compression}, which compresses data while ensuring that it can be reconstructed exactly, and \emph{lossy compression}, which introduces a small \emph{compression error} in order to achieve much higher \emph{compression rates} or \emph{compression factors}, defined here as $\frac{\text{original data size}}{\text{compressed data size}}$. Since our paper is concerned with lossy compression, we will often express this error using one of the following metrics:}

\begin{itemize}
\item The relative error \modified{(RE)} \modified[1]{$\frac{ \| \ten{A}- \widetilde{\ten{A}} \| }{\| \ten{A} \|}$, where $\|\cdot\|$ denotes the Euclidean norm (see \cref{sec:tensor-preliminaries})}
\item The sum-of-squared-errors (SSE) \modified[1]{$\| \ten{A} - \widetilde{\ten{A}} \|^2$.}
\end{itemize}

\modified{When processing numerical data, the floating-point format complicates efficient lossy compression. For example, the sign bit or exponent bits of a floating-point number will have a much larger influence on the represented number than the least significant bits of the mantissa. Therefore \emph{quantization} is required, where each number is approximated within some discrete domain that is easier to represent in a compressed format. These quantized values can then be \emph{encoded}, i.e., stored as a bit stream, using lossless compression techniques such as the ones described in \cref{sec:lossless-compression}.}

\subsection{Lossless compression}
\label{sec:lossless-compression}

When no error on the original data can be tolerated, it can be compressed using \modified[1]{lossless compression}. Many techniques for this purpose were proposed \cite{data-compression}, resulting in general lossless compressors such as zlib \cite{zlib} and Zstandard \cite{zstandard}. Below, we briefly describe two techniques that are relevant to our own work. \modified[1]{Lossless compression can also be used as a component of a lossy compression pipeline.}

\paragraph{Run-length encoding} When storing a sequence of identical symbols, it is more frugal to represent it as a tuple containing this symbol and the length of the sequence, i.e. its \emph{run-length}. For example, this technique is used in JPEG \modified[1]{(Joint Photographic Experts Group)} \cite{jpeg} image compression to compress sequences of zero coefficients.

\paragraph{Entropy coding} In information theory, the \emph{entropy} of an information source represented by a discrete probability distribution $X$ consisting of the probabilities $p_1, \dots, p_n$, is defined as $H(X) = - \sum_i p_i \log_2 p_i$ \cite{shannon}. Furthermore, Shannon's source coding theorem implies that, when encoding a stream of symbols identically and independently distributed according to $X$, one needs at least $H(X)$ bits per symbol on average \cite{shannon}. While this sets an upper limit to compression efficiency within this setting, it does not provide us with a constructive algorithm to approach this bound.

One approach to entropy coding consists of representing each symbol as a particular sequence of bits, its \textit{code word}, and concatenating all code words corresponding to the symbols from the original stream. To allow for unambiguous decoding, the code must be a \emph{prefix code}, i.e., each code word of length $l$ must be different from the first $l$ bits of any other code word. \emph{Huffman coding} \cite{huffman} provides a simple algorithm to produce an optimal prefix code tree and has become widely used since its invention \cite{data-compression}. However, all prefix-code-based methods suffer from the constraint that each code word must consist of an integer number of bits. In contrast, the entropy of a particular symbol, $-\log_2 p_i$, is almost always non-integer. Therefore, this discrepancy decreases the compression efficiency of such methods.

\emph{Arithmetic coding} \cite{arithmetic-coding} solves this issue by representing all symbols in the stream combined as an interval bounded by two long fractions. The compressed output then consists of the binary representation of an arbitrary number within this interval. Since certain bits in this sequence contain information on more than one symbol, the aforementioned limitation of prefix codes is avoided. In fact, when the number of encoded symbols goes to infinity, the theoretical compression efficiency approaches the entropy limit \cite{arithmetic-coding}.

More recently, \emph{asymmetric numeral systems} \cite{ans} were proposed as a faster alternative to arithmetic coding\modified[1]{, while maintaining optimal theoretical compression efficiency.} Because of this speed advantage, asymmetric numeral systems have become widely used in state-of-the-art compressors \cite{zstandard,draco,jpeg-xl}.

\subsection{Lossy compression}
\label{sec:lossy-compression}

\modified[1]{When the original data does not need to be reconstructed exactly, lossy compression can be used.} This is especially relevant for data that is intended for human interpretation, such as images, sound and video. Because of this, many lossy compressors choose which information to discard based on a model of human perception \cite{jpeg,x265}. Alternative error metrics such as the \modified[1]{structural similarity index measure} (SSIM) \cite{ssim} and \modified[1]{video multimethod assessment fusion} (VMAF) \cite{vmaf} were proposed to address this. Nevertheless, in this work we will employ the usual Euclidean (or Frobenius) norm as our error metric, because it is simple and has useful properties with regard to multilinear algebra.

A first class of lossy compressors can be described as \emph{predictive}, in the sense that values are predicted using e.g., neighboring values or a fitting function \cite{compression-survey}. If the predictor performs well, the deviation of each value from its prediction should be small, which makes them easier to encode than the original values. Before encoding, these deviations are quantized in an appropriate way to achieve the desired level of accuracy. For our application, the compression of numerical multi-dimensional arrays, predictive-based compressors include SZ \cite{sz3-1,sz3-2,sz3-3} and FPZIP \cite{fpzip}.

Some compressors apply a certain invertible transformation to the data before quantization. By exploiting the structured nature of most relevant data, these transformations can significantly increase the sparsity of the coefficients in the transformed domain, improving compressibility. The discrete cosine transform (DCT) \cite{dct} is a widely used tool for this purpose \cite{dct-book}, being used in various image and video compressors such as JPEG, x265 \cite{x265} and AV1 \cite{av1}. For our application, transforms are used in compressors such as ZFP \cite{zfp}, TTHRESH \cite{tthresh} and SSEM \cite{ssem}.

Transform-based methods usually rely on fixed transformations, which removes the need to store the transform but might lead to less efficient compression than if a \modified[1]{\emph{data-dependent transform}} were used. For example, when compressing an $n \times n$ \modified[1]{matrix $A$ with singular value decomposition (SVD) $A = U S V^T$, it can be interesting to use its bases of left and right singular vectors $U$ and $V$ as data-dependent transforms applied to the column and row spaces}, respectively, since the Eckart-Young theorem \cite{eckart-young} states that this provides us with the best low-rank approximation in terms of the Euclidean norm. However, even when truncating to rank $r$, the \modified[1]{compressed representation would require $2nr + r$ scalars to store the largest $r$ singular values and the $r$ corresponding columns of $U$ and $V$, compared to $r^2$ when using the same truncation rank for a fixed transform, such as a $2$-dimensional DCT, due to the transform storage cost.}

In the case of data with $3$ or more modes, the \emph{orthogonal Tucker decomposition} \cite{tucker} reduces the relative storage overhead of the transformation by representing an $n_1 \times n_2 \times \dots \times n_d$ tensor as a core tensor of the same size and \modified[1]{orthogonal matrices $U_i \in \mathbb{R}^{n_i \times n_i}$. Like a truncated SVD, we can truncate the matrices $U_i$ to the first $r_i$ columns and retain only the $r_1 \times r_2 \times \cdots \times r_d$ core tensor corresponding to this selection of basis vectors. This reduces the storage cost to $r_1 r_2 \dots r_d$ scalars for the transformed coefficients and $\sum_{i=1}^d n_i r_i$ scalars for the factor matrices.} As such, the number of transformed coefficients grows exponentially in the number of modes $d$, while the transform storage cost only increases linearly with $d$. Due to this relatively low transform storage cost, the Tucker decomposition can still be competitive compared to methods using fixed transforms. In the field of tensor decompositions, this decomposition is also commonly used for compression of tensors of moderate order \cite{kolda-bader,tensor-data-mining}. \modified{For example, the TuckerMPI software efficiently computes Tucker decompositions of massive datasets in a distributed-memory parallel setting.}

\modified[1]{Only few publications discuss the quantization and encoding of such decompositions \cite{tucker-simple-quantization-1,tucker-simple-quantization-2,pre_tthresh,tthresh}}. \modified[1]{To our knowledge, t}he Tucker-based TTHRESH compressor \modified[1]{employs the most advanced quantization scheme among these,} using \emph{bit plane truncation}, which will be discussed further in \cref{sec:bit-plane-truncation}. We will use a variant of this strategy in ATC.

Tensor train \cite{tensor-trains} and hierarchical Tucker \cite{hierarchical-tucker} decompositions are also used for compression of high-order tensors \cite{tensor-data-mining}. 
In our targeted application we are dealing with dense tensors, i.e., tensors which store each element explicitly, so the order of tensor is typically not high. Therefore, these alternative decompositions will not be discussed further.

\modified[4]{Finally, a new category of low-rank tensor formats based on subpartitioning is also emerging \cite{adaptive-subtensor-partitioning,multiresolution-low-rank-tensor-formats}. Many data compressors already split the input data into subblocks before compression \cite{jpeg,sz3-1,zfp,h264}, not only to reduce the complexity of most transform-based methods, but also to improve compression when adaptively partitioning the data into blocks to diminish intra-block discontinuities. While it is therefore promising to apply this approach to tensor formats, this is out of scope for the current version of ATC.}
\section{Preliminaries}
\label{sec:preliminaries}


\subsection{Tensor concepts and notation}
\label{sec:tensor-preliminaries}

\sloppy A tensor \modified[1]{$\ten{A} \in \mathbb{R}^{n_1 \times \dots \times n_d}$} is represented in bases by a multidimensional array containing numerical values indexed with $d$ integers. \modified[1]{$n_1 \times \dots \times n_d$ is called the size of the tensor, and $d$ is called the order.} The \emph{Euclidean inner product} of two tensors $\langle \ten{X}, \ten{Y} \rangle$ is the sum over their elementwise products. The induced \emph{Euclidean norm} is $\norm{\ten{X}} = \sqrt{\langle \ten{X}, \ten{X} \rangle}$.

\modified[1]{We will use the Matlab notation $a:b$ as a tensor index to select a subrange along the corresponding mode from index $a$ up to and including index $b$, or simply $:$ in case all indices are selected. For example, $U_{:,1:r}$ refers to the submatrix containing all rows but only the first $r$ columns of $U$.} A \emph{mode-$k$ fiber} of a tensor $\ten{X}$ is then defined as a vector obtained by fixing all indices in the tensor apart from the $k$-th index, i.e. $\ten{X}_{i_1, \dots, i_{k-1}, :, i_{k+1}, \dots, i_d} \in \mathbb{R}^{n_k}$. Conversely, a \emph{mode-$k$ slice} is a tensor of order $d - 1$ obtained by only fixing the $k$-th index of the tensor, i.e $\ten{X}_{:, \dots, :, i_k, :, \dots, :} \in \mathbb{R}^{n_1 \times \dots \times n_{k-1} \times n_{k+1} \times \dots \times n_d}$.

By arranging all mode-$k$ fibers as columns in a matrix in a consistent manner, we obtain the mode-$k$ matricization of $\ten{X}$, denoted by \modified[1]{$X_{(k)} \in \mathbb{R}^{n_k \times \Pi_{i \neq k} n_i}$. We then define the \emph{mode-$k$ matrix-tensor product} as follows:}
\[
\ten{Y} = U \mtprod{k} \ten{X} \Leftrightarrow Y_{(k)} = U X_{(k)} .
\]
In other words, multiplying a tensor by a matrix $U$ along mode $k$ is equivalent with transforming each mode-$k$ fiber using $U$. This operation has several useful properties:
\begin{itemize}
\item The multilinear product commutes along different modes, i.e., $U \mtprod{i} U' \mtprod{j} \ten{X} = U' \mtprod{j} U \mtprod{i} \ten{X}$ for $i \neq j$.
\item Multilinear products along the same mode can be composed, i.e., $U \mtprod{i} U' \mtprod{i} \ten{X} = (U U') \mtprod{i} \ten{X}$.
\item Because multiplying a matrix by an orthogonal matrix preserves its Euclidean norm, multiplying a tensor by an orthogonal matrix along any mode also preserves its Euclidean norm.
\end{itemize}

\subsection{ST-HOSVD rank truncation}
Using the mode-$k$ product, we define a Tucker decomposition \cite{tucker} of $\ten{A} \in \mathbb{R}^{n_1 \times \dots \times n_d}$ as
\[
(U_1, \dots, U_d) \tuckerprod \ten{B} = U_1 \mtprod{1} \,\cdots\, \mtprod{d-1} U_d \mtprod{d} \ten{B} ,
\]
where $\ten{B} \in \mathbb{R}^{n_1 \times \dots \times n_d}$ is called the core tensor and $\modified[1]{(U_1, \dots, U_d) \in \mathbb{R}^{n_1 \times n_1} \times \dots \times \mathbb{R}^{n_d \times n_d}}$ are called the factor matrices. A common method to compute a Tucker decomposition of a tensor $\ten{A}$ is the \emph{higher-order singular value decomposition} (HOSVD) \cite{hosvd}, where each \modified[1]{orthogonal factor matrix} $U_i$ is chosen as the matrix of left singular vectors of $A_{(i)}$. A truncated HOSVD can then be obtained by \modified[1]{only retaining the first $r_i \leqslant n_i$ columns}, leading to a truncated core $\overline{\ten{B}} \in \mathbb{R}^{r_1 \times \dots \times r_d}$ and factors $\overline{U_i} \in \mathbb{R}^{n_i \times r_i}$. Alternatively, the  ST-HOSVD \cite{st_hosvd} algorithm can also be used, which interleaves the factor computation steps with the rank truncation and projection steps. The full procedure is shown in \cref{alg:st-hosvd}. Due to the data reduction in each iteration, this method significantly speeds up as more truncation is applied. Furthermore, the resulting compression error is almost always less than or equal to the one obtained by the truncated HOSVD \cite{st_hosvd}.

\begin{algorithm}[t]

\setstretch{0.9}
\SetAlgoLined
\KwData{input tensor $\ten{A}$ of order $d$}
\KwResult{truncated core $\overline{\ten{B}}$, truncated factors $\overline{U_1}, \dots, \overline{U_d}$}

$\overline{\ten{B}} = \ten{A}$\;
\For{$i = 1, \dots, d$}{
 \modified[1]{[Compute a singular value decomposition $\overline{B}_{(i)} = U \Sigma V^T$]}\;
 [Choose truncation rank $r_i$]\;
 $\overline{U_i} = U_{:, 1:r_i}$\;
 $\overline{B}_{(i)} = \Sigma_{1:r_i,1:r_i} V_{:,1:r_i}^T$\;
}

\caption{The ST-HOSVD \cite{st_hosvd}}
\label{alg:st-hosvd}
\end{algorithm}

For the remainder of this paper, we assume the original data to be compressed is an order-$d$ tensor $\ten{A} \in \mathbb{R}^{n_1 \times n_2 \times \dots \times n_d}$ with total size $N = \Pi_{i=1}^d n_i$. This tensor will be approximated by the multilinear rank $(r_1, \dots, r_d)$ ST-HOSVD $(\overline{U_1}, \dots, \overline{U_d}) \tuckerprod \overline{\ten{B}}$, \modified[1]{where $\modified[4]{\overline{\ten{B}}}\in\mathbb{R}^{r_1\times\dots\times r_d}$ and the $\overline{U}_i \in \mathbb{R}^{n_i \times r_i}$ have orthonormal columns.} The final approximation $\widetilde{\ten{A}}$ produced by the proposed ATC compressor will be denoted by $(\widetilde{U_1}, \dots, \widetilde{U_d}) \tuckerprod \widetilde{\ten{B}}$, where $\widetilde{\ten{B}} \in \mathbb{R}^{r_1 \times \dots \times r_d}$ and $\widetilde{U_i} \in \mathbb{R}^{n_i \times r_i}$.

\subsection{TTHRESH bit plane truncation and encoding}
\label{sec:bit-plane-truncation}

While compression can be achieved through the aforementioned concept of rank truncation, another strategy is to compute the full HOSVD $\ten{A} = (U_1, \dots, U_d) \tuckerprod \ten{B}$ and then store the resulting coefficients with limited precision. This is the essence of the bit plane truncation scheme employed by the Tucker-based compressor TTHRESH \cite{tthresh}, which then encodes the remaining data using a custom procedure described below. \Cref{tab:bit-plane-truncation} demonstrates this process with an example \modified[1]{where $\ten{B} \in \mathbb{R}^{2 \times 2 \times 2}$}.

\begin{table}[t]
\setlength{\tabcolsep}{4pt}
\centering
\begin{tabular}{l|c|lllllllllll}
          & \multicolumn{1}{l|}{Sign}                        & \multicolumn{11}{l}{Absolute value (binary)}                                                                                                                                                                                                                                                                                                                                                                                                              \\ \hline
$\modified[1]{b}_{111}$ & {\bf 0}                         & {\it 1} & {\bf 0} & {\bf 0} & {\bf 0} & {\bf 0} & {\bf 0}                      & \multicolumn{1}{l|}{{\bf 0}}    & {\color{lightgray} 1} & {\color{lightgray} 1} & {\color{lightgray} 1} & {\color{lightgray} $\dots$} \\
$\modified[1]{b}_{112}$ & {\color{lightgray} 0} & {\it 0} & {\it 0} & {\it 0} & {\it 0} & {\it 0} & {\it 0}                      & \multicolumn{1}{l|}{{\it 0}}    & {\color{lightgray} 0} & {\color{lightgray} 1} & {\color{lightgray} 1} & {\color{lightgray} $\dots$} \\
$\modified[1]{b}_{121}$ & {\color{lightgray} 0} & {\it 0} & {\it 0} & {\it 0} & {\it 0} & {\it 0} & {\it 0}                      & \multicolumn{1}{l|}{{\it 0}}    & {\color{lightgray} 0} & {\color{lightgray} 0} & {\color{lightgray} 0} & {\color{lightgray} $\dots$} \\
$\modified[1]{b}_{122}$ & {\bf 1}                         & {\it 0} & {\it 0} & {\it 0} & {\it 1} & {\bf 1} & {\bf 0}                      & \multicolumn{1}{l|}{{\bf 0}}    & {\color{lightgray} 1} & {\color{lightgray} 0} & {\color{lightgray} 1} & {\color{lightgray} $\dots$} \\ \cline{9-9}
$\modified[1]{b}_{211}$ & {\color{lightgray} 0} & {\it 0} & {\it 0} & {\it 0} & {\it 0} & {\it 0} & \multicolumn{1}{l|}{{\it 0}} & {\color{lightgray} 1} & {\color{lightgray} 0} & {\color{lightgray} 0} & {\color{lightgray} 1} & {\color{lightgray} $\dots$} \\
$\modified[1]{b}_{212}$ & {\bf 1}                         & {\it 0} & {\it 0} & {\it 0} & {\it 1} & {\bf 0} & \multicolumn{1}{l|}{{\bf 0}} & {\color{lightgray} 0} & {\color{lightgray} 1} & {\color{lightgray} 1} & {\color{lightgray} 0} & {\color{lightgray} $\dots$} \\
$\modified[1]{b}_{221}$ & {\bf 0}                         & {\it 0} & {\it 0} & {\it 0} & {\it 1} & {\bf 0} & \multicolumn{1}{l|}{{\bf 0}} & {\color{lightgray} 1} & {\color{lightgray} 1} & {\color{lightgray} 0} & {\color{lightgray} 0} & {\color{lightgray} $\dots$} \\
$\modified[1]{b}_{222}$ & {\bf 0}                         & {\it 0} & {\it 0} & {\it 0} & {\it 0} & {\it 1} & \multicolumn{1}{l|}{{\bf 0}} & {\color{lightgray} 1} & {\color{lightgray} 0} & {\color{lightgray} 0} & {\color{lightgray} 1} & {\color{lightgray} $\dots$}
\end{tabular}
\caption{\modified[4]{Example of TTHRESH-style bit plane truncation of an example $2 \times 2 \times 2$ core and the different bit categories that remain.}}
\label{tab:bit-plane-truncation}
\end{table}

First, all entries of $\ten{B}$ are scaled by \modified[1]{$2^k$ where $k \in \mathbb{N}$ such that $2^{63} \leqslant 2^k \|\ten{B}\|_{\max} < 2^{64}$, where $\|\cdot\|_{\max}$ denotes the max-norm: the largest absolute value of the entries in the tensor}. This scaling factor is stored in the compressed output so the process can be inverted during decompression. Then, the core coefficients of $2^k \ten{B}$ are rounded to the nearest integer. By vectorizing the core coefficients into \modified[1]{a column vector of length $\Pi_i r_i$} and considering the binary representation of each coefficient's absolute value, we obtain a bit matrix, in which each row represents a single coefficient. \modified[1]{The first few columns of this matrix are shown in \cref{tab:bit-plane-truncation}.} We then iterate over each column of this matrix, i.e., each \emph{bit plane}, starting from the left, i.e., in order of significance. Within each bit plane, we process each bit one by one and track the core quantization error that would be achieved by encoding all bits up till this point. Due to the orthogonality of the Tucker factors, ignoring perturbations introduced during factor compression, this core error will equal the final compression error. Therefore, when the target compression error is reached, the procedure ends. We define the point in the bit matrix where this happens as the \emph{breakpoint} of the bit plane truncation process. \modified[1]{In \cref{tab:bit-plane-truncation}, the breakpoint is at position 4 on the $7$th-highest bit plane. Furthermore, we define all coefficients with at least one encoded 1-bit as \emph{significant}, with all other coefficients being \emph{insignificant}. In \cref{tab:bit-plane-truncation}, coefficients $b_{112}$, $b_{121}$ and $b_{211}$ are insignificant, while all others are significant.}

All bits \modified[1]{in the vectorized core} up to the breakpoint are included in the encoded core. Not all bits are encoded in the same way.
The \emph{leading bits} of each coefficient, marked in \modified[4]{italics} in \cref{tab:bit-plane-truncation}, mostly consist of zeroes. Therefore, run-length encoding is applied to all sequences of zeroes within each bit plane\modified[1]{, i.e., within each column of the bit matrix}. Because the resulting run-lengths are highly non-uniformly distributed \cite{tthresh}, they are then compressed further using arithmetic coding.
The \textit{sign bits} of the significant coefficients as well as the \emph{trailing bits} (marked in \modified[4]{bold}) are almost uniformly distributed in general, so they are encoded without compression.

Note that all insignificant coefficients will be decoded as zero, so their signs do not need to be stored. Furthermore, it is possible that after bit plane truncation, certain core slices only contain insignificant coefficients, which are represented by 0 in the quantized core. Therefore, the corresponding factor columns will not be encoded. For further details, such as the quantization and encoding of the factors, we refer the reader to \cite{tthresh}.

\modified[1]{Ballester-Ripoll and Pajarola \cite{pre_tthresh} concluded} that a variant of the aforementioned thresholding scheme consistently leads to better compression than Tucker rank truncation followed by a simple quantization scheme. After all, rank truncation indiscriminately removes all coefficients from low-energy slices, while bit plane truncation preserves the most significant components from all slices. We will use a variation of this approach in ATC.

\section{The ATC pipeline}
\label{sec:algorithmic_improvements}

\begin{figure}[t]
    \centering
    \includegraphics[width=.9\textwidth]{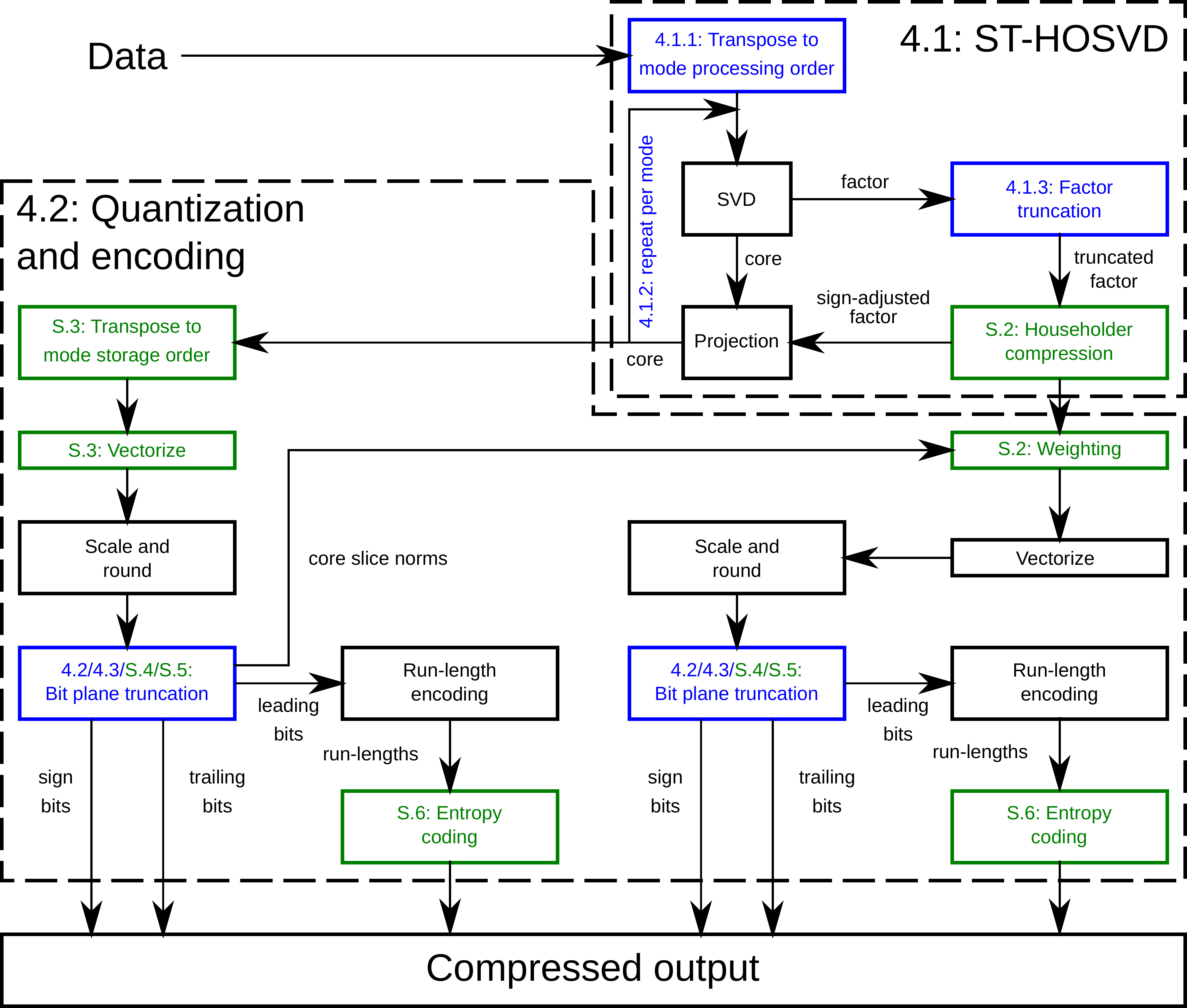}
    \caption{\modified{Overview of the ATC compression pipeline. The \modified[5]{blue} components \modified[5]{(with section labels 4.1.1, 4.1.2, 4.1.3, 4.2 and 4.3)} indicate \modified[5]{the most important} parts of the pipeline that were added or modified with respect to TTHRESH and are described in the corresponding sections. The green components \modified[4]{(with section labels S.2, S.3, S.4, S.5 and S.6)} only represent minor performance improvements and are therefore described in the supplementary material.}}
    \label{fig:flowchart}
\end{figure}

In this section we describe the ATC pipeline, which is summarized in \cref{fig:flowchart}. This diagram serves as a point of reference for the reader throughout this section. While the decompression pipeline also has a few distinctive components, it is very similar to the compression procedure in reverse. Therefore, it is not documented in a separate diagram.

In the first phase of the compression process, the data is processed by an ST-HOSVD. We describe the motivation for ST-HOSVD rank truncation and the implications for error control when combined with bit plane truncation in \cref{sec:hybrid_truncation}. During the second phase, the resulting core tensor and factors are quantized and encoded. Although this is based on TTHRESH's bit plane truncation scheme, we applied several improvements, two of which will be discussed in \cref{sec:parallel-quantization-and-encoding,sec:dequantization-correction-prediction}. \modified{Various minor improvements are discussed in the supplementary material.}

\subsection{\modified{Hybrid truncation}}
\label{sec:hybrid_truncation}

As discussed in \cref{sec:preliminaries}, previous research suggests that while the ST-HOSVD can compute truncated Tucker decompositions relatively quickly, TTHRESH's bit plane truncation approach achieves superior compression. Combining both strategies was suggested by \cite{pre_tthresh,tthresh} as a trade-off between compression rate and execution time.

In our approach, we first apply rank truncation during the ST-HOSVD (see \cref{alg:st-hosvd}) to approximate a certain target rank truncation error. Then, the truncated core and factors are processed during the quantization and encoding phase as in TTHRESH, introducing a quantization error as well. 

\subsubsection{ST-HOSVD mode processing order}
Due to the truncation performed in each step of the ST-HOSVD, the mode processing order can significantly influence execution time. In ATC, the modes are processed in order of increasing mode length during compression by default, following the heuristic described in \cite{st_hosvd}. However, ATC also allows the user to specify a processing order. For example, if a particular mode is known to be highly compressible a priori, it is advisable to process it first, since this will lead to a large reduction of the core size for the subsequent steps.

During decompression, the mode processing order $p$ is determined by minimizing the approximate number of operations required, estimated as $\sum_{i=1}^d n_{p_1} \dots n_{p_i} r_{p_i} \dots r_{p_d}$ based on the complexity of naive matrix multiplication. This optimal order is found through exhaustive search, which takes $O(d d!)$ time. While this complexity could be reduced to $O(d \log d)$ as described in section 8 of \cite{tuckermpi}, we did not implement such an algorithm due to the very low value of $d$.

\subsubsection{Circular mode shift}
\label{sec:circular-mode-shift}
To minimize the number of tensor transpositions needed in the ST-HOSVD, we employ a \emph{circular mode shift} trick, which lets us compute the ST-HOSVD with arbitrary mode processing order using only a single transposition, regardless of the order of the tensor. \Cref{alg:circular-mode-shift} describes the full procedure. We start by transposing the input tensor, such that the core is initially stored with its modes ordered in the same way as the mode processing order $p$. In the first iteration, we matricize the tensor by ``merging'' all but the first mode, resulting in an $n_{p_1} \times n_{p_2} n_{p_3} \dots n_{p_d}$ matrix (line 3). Mode $p_1$ can then be processed. After selecting an appropriate factor matrix, we project the core while transposing it, which results in an $n_{p_2} \dots n_{p_d} \times r_{p_1}$ matrix (line 5). The matrix can now be reshaped into a tensor with mode order $(p_2, \dots, p_d, p_1)$ (line 6). Therefore, each iteration effectively applies a circular shift to the mode order. By initially transposing this mode order to the order $p$, we ensure that at the start of the $i$-th step, the mode $p_i$ will be at the front of the mode order, allowing it to be processed.

\begin{algorithm}[t]
\setstretch{0.9}

\SetAlgoLined
\KwData{input tensor $\ten{A}$ stored in the default mode order $(1, \dots, d)$, mode processing order $p$}
\KwResult{truncated core $\overline{\ten{B}}$ stored in the mode order $(p_1, \dots, p_d)$, truncated factors $\overline{U_1}, \dots, \overline{U_d}$}

$\overline{\ten{B}} = \text{transpose}(\ten{A}, p)$\;
\For{$i = 1, \dots, d$}{
 $B = \text{reshape}(\overline{\ten{B}}, (n_{p_i}, n_{p_{i+1}} \dots n_{p_d} r_{p_1} \dots r_{p_{i-1}}))$\;
 [Determine factor $U_{p_i} \in \mathbb{R}^{n_{p_i} \times r_{p_i}}$ based on the singular value decomposition of $B$]\;
 $\overline{B} = B^T U_{p_i}$\;
 $\overline{\ten{B}} = \text{reshape}(\overline{B}, (n_{p_{i+1}}, \dots, n_{p_d}, r_{p_1}, \dots, r_{p_i}))$\;
}

\caption{ST-HOSVD compression with circular mode shift}
\label{alg:circular-mode-shift}

\end{algorithm}

Note that \modified[4]{these} reshaping operations simply reinterpret \modified[4]{the same memory as matrices with different dimensions and do not result in any data movement or poor memory access patterns, leading to} a negligible runtime cost. Furthermore, \modified[4]{the Eigen linear algebra software library \cite{eigen}, used in ATC,} evaluates the transposition and matrix multiplication on line 5 simultaneously using expression templates, so effectively no transposition needs to be processed apart from the initial one on line 1. We implemented a similar procedure for the decompression process as well.

\subsubsection{Error control}
The target error for ST-HOSVD is determined by the parameter \verb|rank_truncation_max_sse_share| (RTMSS) \modified{as follows:}
\[
\modified{\modified[1]{\text{TargetSSE}_\text{ST-HOSVD} = \text{RTMSS} \cdot \text{TargetSSE}_\text{total}, \quad \text{where} \quad 0 \leqslant \text{RTMSS} \leqslant 1.}}
\]
Using this target, the truncation rank for each mode is determined dynamically using the strategy described in \cite[section 6.3]{st_hosvd}. The actual rank truncation SSE can simply be computed as the sum of the squares of the singular values discarded across all steps of the ST-HOSVD, as stated in theorem 6.4 from \cite{st_hosvd}.

To control the final error of the decompressed tensor, we will now approximately separate this error into multiple components. \modified[1]{Based on the ST-HOSVD $\ten{A} \approx (\overline{U_1}, \dots, \overline{U_d}) \tuckerprod \overline{\ten{B}}$, we define the full factors $U_i$ as arbitrary orthogonal matrices such that $(U_i)_{:,1:r_i} = \modified{\overline{U_i}}$, with the corresponding full core $\ten{B} = (U_1^T, \dots, U_d^T) \tuckerprod \ten{A}$. To simplify arithmetic, we also introduce the padded truncated factors $\modified{\overline{U_i}'} \in \mathbb{R}^{n_i \times n_i}$ with $(\modified{\overline{U_i}'})_{:,1:r_i} = \overline{U_i}$ and $(\modified{\overline{U_i}'})_{:,r_i + 1:n_i} = 0$.} \modified[1]{We define the padded quantized core ${\widetilde{\ten{B}}'} \in \mathbb{R}^{n_1 \times \dots \times n_d}$ as a tensor containing the quantized core coefficients in ${\widetilde{\ten{B}}'}_{1:r_1, \dots, 1:r_d}$ and $0$ everywhere else. This tensor will be decomposed as the sum of the full core $\ten{B}$, the truncation error tensor $\overline{\ten{B}}' - \ten{B}$ and the quantization error tensor $\modified[4]{\delta \ten{B} = \widetilde{\ten{B}}' - \overline{\ten{B}}'}$. The quantized factors ${\widetilde{U_i}'} \in \mathbb{R}^{n_i \times n_i}$ are matrices with the} first $r_i$ columns consisting of the quantized factor coefficients and $0$ elsewhere. Furthermore, we define $\delta U_i = {\widetilde{U_i}'} - {\overline{U_i}'}$. Note that due to the non-zero pattern in ${\widetilde{\ten{B}}'}$ this means that ${\overline{U_i}'} \mtprod{i} {\widetilde{\ten{B}}'} = U_i \mtprod{i} {\widetilde{\ten{B}}'}$ for $i = 1, \dots, d$. \modified[1]{Recall that $\widetilde{\ten{A}}$ is the final approximation.} We start with the following expression:
\[
\norm{\widetilde{\ten{A}} - \ten{A}}^2
= \norm{({\overline{U_1}'} + \delta U_1, \dots, {\overline{U_d}'} + \delta U_d) \tuckerprod {\widetilde{\ten{B}}'} - (U_1, \dots, U_d) \tuckerprod \ten{B}}^2 .
\]
Discarding higher-order factor quantization errors we get:
\begin{align}
\norm{\widetilde{\ten{A}} - \ten{A}}^2
&\approx \norm{(\overline{U_1}, \dots, \overline{U_d}) \tuckerprod {\widetilde{\ten{B}}'} + \sum_{i=1}^d ({\overline{U_1}'}, \dots, {\overline{U_{i-1}}'}, \delta U_i, {\overline{U_{i+1}}'}, \dots, {\overline{U_d}'}) \tuckerprod {\widetilde{\ten{B}}'} - (U_1, \dots, U_d) \tuckerprod \ten{B}}^2 \nonumber \\
&= \norm{{\widetilde{\ten{B}}'} - \ten{B} + \sum_{i=1}^d (U_i^T \delta U_i) \mtprod{i} {\widetilde{\ten{B}}}'}^2 \nonumber \\
&= \norm{({\overline{\ten{B}}'} - \ten{B}) + \delta \ten{B} + \sum_{i=1}^d (U_i^T \delta U_i) \mtprod{i} {\widetilde{\ten{B}}'}}^2 \nonumber \\
&= \norm{{\overline{\ten{B}}'} - \ten{B}}^2 + \norm{\delta \ten{B}}^2 + \sum_{i=1}^d \norm{(U_i^T \delta U_i) \mtprod{i} {\widetilde{\ten{B}}'}}^2 \nonumber \\
&\hspace{3cm}+2 \sum_{i=1}^d \langle {\widetilde{\ten{B}}'} - \ten{B}, (U_i^T \delta U_i) \mtprod{i} {\widetilde{\ten{B}}'} \rangle + 2 \sum_{i=1}^d \sum_{j=1}^{i-1} \langle (U_i^T \delta U_i) \mtprod{i} {\widetilde{\ten{B}}'}, (U_j^T \delta U_j) \mtprod{j} {\widetilde{\ten{B}}'} \rangle ,
\label{eq:error-approximation-precise}
\end{align}
where we used that the truncation and quantization error tensors $({\overline{\ten{B}}'} - \ten{B})$ and $\delta \ten{B}$ have complimentary non-zero patterns, so their inner product is zero. Using the Cauchy-Schwartz inequality, we can provide an upper bound:
\[
\norm{\widetilde{\ten{A}} - \ten{A}}^2 \lessapprox \norm{{\overline{\ten{B}}'} - \ten{B}}^2 + \norm{\delta \ten{B}}^2 + \sum_{i=1}^d \norm{\delta U_i \mtprod{i} {\widetilde{\ten{B}}'}}^2 + 2 \sum_{i=1}^d \norm{\delta U_i \mtprod{i} {\widetilde{\ten{B}}'}} \left( \norm{{\widetilde{\ten{B}}'} - \ten{B}} + \sum_{j=1}^{i-1} \norm{\delta U_j \mtprod{j} {\widetilde{\ten{B}}'}} \right).
\]
However, \modified{for the general, dense tensors in our setting,} this bound is very loose in practice. This can be explained by considering that due to the mostly uniform quantization errors, the error tensor components are not aligned in a particular direction. Because these tensors live in a very high-dimensional space, they are almost orthogonal to each other. Thus, the inner products in \cref{eq:error-approximation-precise} are dominated by the squared norms in practice.

As a result, we propose to ignore the inner products from \cref{eq:error-approximation-precise} to obtain a more useful estimate for the compression error. Furthermore, note that the core tensor $\ten{C}$ produced by an HOSVD is \emph{all-orthogonal}, i.e., all mode-$i$ slices are orthogonal to each other for each $i$ \cite{hosvd}, which implies that $\norm{\delta U_i \mtprod{i} \ten{C}} = \norm{\delta U_i \Sigma_i}_F$, where $\Sigma_i$ is a diagonal matrix containing the mode-$i$ core slice norms of $\ten{C}$. Although this property does not hold exactly for ${\widetilde{\ten{B}}'}$ due to truncation and quantization errors, we observed that it is mostly all-orthogonal in practice. Therefore we approximate $\norm{\delta U_i \mtprod{i} {\widetilde{\ten{B}}'}}$ by $\norm{\delta U_i {\widetilde{\Sigma_i}'}}_F$. This leads to the following error approximation:
\begin{equation}
\norm{\widetilde{\ten{A}} - \ten{A}}^2
\approx \norm{{\overline{\ten{B}}'} - \ten{B}}^2 + \norm{\delta \ten{B}}^2 + \sum_{i=1}^d \norm{\delta U_i {\widetilde{\Sigma_i}'}}_F^2 .
\label{eq:total_error_approximation}
\end{equation}

When compressing several datasets \modified[1]{from \cref{sec:datasets} with various RTMSS values and target relative errors $10^{-2}$ and $10^{-3}$}, we found that the \modified[3]{relative difference in between the left-hand and right-hand side of \cref{eq:total_error_approximation} never exceeded \modified[4]{$0.15\%$}. At a target relative error of $10^{-1}$, this deviation increased to $8.5\%$. We conclude that the approximation is usually very accurate,} so we will use it to predict the total error. Finally, like in TTHRESH we do not consider the factor errors while quantizing the core because they are not yet known. Therefore we simply set the target core quantization SSE to \modified[1]{$\text{TargetSSE}_{\text{total}} - \text{ActualSSE}_{\text{ST-HOSVD}}$}.

\begin{figure}[t]
	\centering
	\input{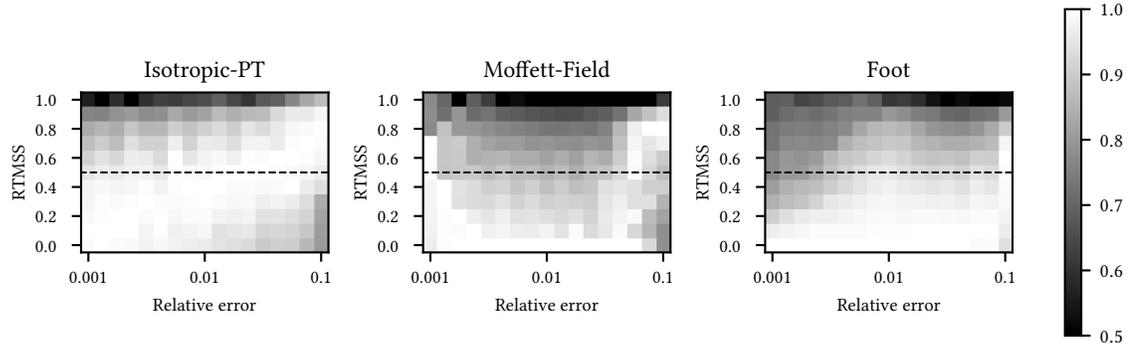}
	\vspace{-3em}
	\caption{\modified[4]{Relative compression factors across different errors and datasets in terms of RTMSS. Each compression factor is normalized by dividing it by the maximum compression factor for the same error. The dashed line indicates the default RTMSS value, which balances suboptimal compression rates with increased \modified[1]{compression and decompression} speed.}}
	\label{fig:rtmss-rate-distortion}
\end{figure}

\Cref{fig:rtmss-rate-distortion} shows the effect of the RTMSS parameter on the compression factor, with $0$ and $1$ corresponding to almost no and almost only rank truncation respectively. Surprisingly, some level of hybrid truncation improves the compression rate in certain cases, in addition to improving the execution time, although the optimal RTMSS parameter depends on the dataset and compression error. This improvement can be attributed to a very large reduction in encoded leading zeroes, which generally also decreases their cost after run-length and entropy coding. We chose $0.5$ as the default RTMSS parameter value, which we believe to be a reasonable trade-off between compression efficiency and speed.

\subsection{Parallel core quantization and encoding}
\label{sec:parallel-quantization-and-encoding}

Since the core quantization and encoding process represents a significant share of the total compression time, we adapted it to support multi-threading. This is achieved by splitting the vectorized core into a series of equally sized blocks, with each block containing a contiguous series of quantized core coefficients. The number of blocks is at least as large as the number of threads. \modified{Each thread can then process and encode the bits of the current bit plane inside one or more blocks independently of the other blocks.} This procedure is executed sequentially for all bit planes until the encoding breakpoint is reached.

\modified{A minor complication is dynamically selecting a breakpoint for approximating the desired quantization error in parallel. To \modified[1]{avoid} synchronization overhead\modified[1]{, we process} bit planes in full without synchronization and check the total error reduction achieved at the end of the bit plane. If this exceeds the target error reduction, we process the bit plane again with a limited number of threads and \modified[3]{synchronized error checks}. When one thread determines that the target error has been reached, all threads will stop encoding bits, leading to a separate breakpoint for each thread. Therefore, the full bit plane will consist of several alternating sections of encoded and non-encoded bits. \modified[4]{Note that only the last encoded bit plane needs to be processed twice, as the global target error reduction can only be exceeded once.}}

\modified{For each block, the bits of the current bit plane are then encoded independently and written to temporary buffers in-memory. Afterwards these compressed blocks are sequentially written to the compressed output, including the size of each block.} Although these sizes could be determined during decompression, explicitly storing them allows ATC to quickly read and separate all blocks into temporary buffers during decompression before parsing them in parallel, leading to parallelized dequantization and decoding as well.

\subsection{Improving error control by predicting the dequantization correction}
\label{sec:dequantization-correction-prediction}

During dequantization, if a decoded coefficient $\tilde{a}$ contains bits down to bit plane $p$, we know that the original value $a$ was located in the interval $[\tilde{a}, \tilde{a} + 2^p - 1]$ (ignoring signs). To decrease the quantization error, we therefore approximate $a$ as $\tilde{a} + 2^{p - 1}$. \modified{While this correction was already applied in TTHRESH, we significantly improved error control by considering this during the quantization error tracking process as well.}

\modified[1]{To demonstrate the significance of this correction, assume that $a$ is drawn from the uniform distribution $U[\tilde{a}, \tilde{a} + 2^p]$. Then, we have that 
\[
 \mathbb{E}_{a\sim U[\tilde{a}, \tilde{a}+2^p]}\left[(a - \tilde{a})^2\right] = \frac{2^{2p}}{3} \quad\text{and}\quad 
 \mathbb{E}_{a\sim U[\tilde{a}, \tilde{a}+2^p]}\left[(a - \tilde{a}-2^{p-1})^2\right] = \frac{2^{2p}}{12}.
\]
Therefore, if we guess that $a$ is (approximately) in the middle of the uncertainty interval rather than at its lower edge, we reduce the SSE by a factor of $4$ when $p$ is large. While the values $a$ are not exactly uniformly distributed in practice, this nevertheless shows that dequantization correction has a significant impact on the final error and therefore needs to be considered during quantization to achieve precise error control.}

\section{Software implementation}
\label{sec:implementation}

ATC is implemented in C++17 and is mainly designed to optimize compression efficiency, speed, error control and memory usage. Furthermore, certain mathematical software aspects, such as the usability of the library, were taken into account. \modified{In this section, we discuss the handling of cut-outs, the interfaces of the library, shared-memory parallelism, libraries, support for various types, I/O, and general software design considerations.} When referring to library version numbers throughout this section, we will specify the lowest tested version.

\paragraph{Cut-outs and downsampling}
In some settings, only a part of the decompressed tensor may be needed. In such situations, the Tucker decomposition can be efficiently combined with the extraction of a \emph{cut-out} from the decompressed tensor. Although any linear filter could be used to downscale the granularity of the tensor grid, we simply decided to reuse the downsampling, box and Lanczos filtering methods from TTHRESH, as described in section 5 of \cite{tthresh}. ATC is implemented in such a way that new filters could be added relatively easily if necessary.

\paragraph{Interfaces}
ATC's native interface is written in C++. However, due to the use of the Standard Template Library (STL) containers this interface may cause compatibility problems when the library and user code are not compiled in exactly the same setting. To address this we provide a C wrapper, which also serves as an interface to C users as well as users which may want to connect ATC to other languages, such as Python. Finally, we implemented an executable which offers a command-line interface too.

\paragraph{Shared-memory parallelism}
\label{sec:parallellism}
The user can specify the \verb|threads| parameter to determine how many threads can be used for linear algebra and tensor transpositions, as these steps are performed using external libraries which support \modified[1]{shared-memory parallelism}. Furthermore, the core quantization and encoding process is parallelized using OpenMP 4.5 \cite{openmp}, as discussed in \cref{sec:parallel-quantization-and-encoding}, and can be controlled using the same parameter.

\paragraph{Libraries used}
Tensor transpositions are performed using the High-Performance Tensor Transpose (HPTT) library version 1.0.5 \cite{hptt}. This library supports arbitrary mode permutations, allowing ATC to use a custom mode processing order during the ST-HOSVD, as well as a custom mode storage order for quantization and encoding. We observed that HPTT appreciably sped up these parts of the compression pipeline. Some code from TTHRESH \cite{tthresh} was reused with permission from the author to implement certain shared features, such as the handling of cut-outs.

Linear algebra is performed using \modified[4]{Eigen 3.3.4}, \modified[1]{with an option to use Basic Linear Algebra Subprograms (BLAS) \cite{blas} for matrix multiplications}. \modified{To enable an optional higher-order DCT preprocessing phase, the Fastest Fourier Transform in the West (FFTW) 3 library \cite{fftw} is needed.} If the user wishes to compile the ATC command-line interface, the Boost.Program\_options library (version 2) \cite{boost-program-options} is also required. Finally, to enable multi-threading in some parts of the pipeline as discussed \modified[3]{in the previous paragraph}, OpenMP 4.5 \cite{openmp} is required.

\paragraph{Supported types and I/O}
\label{sec:supported-types}
ATC supports a wide range of data types for the input tensor: $8$, $16$, $32$ and $64$-bit integers, signed and unsigned, as well as $32$ and $64$-bit floating-point numbers. The tensor can be provided either as a file or as a buffer in memory. Similarly, both the compressed and decompressed output can be stored using either method. The original and decompressed tensors are stored in a flattened, binary format. Metadata such as the data type and mode sizes are passed as separate arguments during compression and are stored in the compressed file for use during decompression. If the original data file contains a header, this can be ignored using the optional \verb|skip_bytes| parameter.

Internally, floating-point arithmetic is performed using $64$-bit precision by default, although the user can switch to $32$-bit precision if desired. We provide a similar option for choosing in between $32$ and $64$-bit integer types for quantization and encoding. Our experiments indicate that in some high-error cases, lower precision suffices to achieve roughly the same compression factor while significantly improving speed.

\paragraph{Software design}
To support the various types described in \modified[3]{previous paragraph}, ATC uses C++ templates throughout most of its code. While this keeps the implementation concise and improves modifiability, by default templated code must be written in header files so the compiler can instantiate the code only for the requested types when compiling the user code. Therefore the library templated code needs to be recompiled by the user whenever the application is modified, which can lead to long compilation times and can slow down development. Since ATC only supports a fixed set of types, we addressed this issue by explicitly instantiating the relevant code for each of the valid types for most type parameters.

However, instantiating all code in terms of each type parameter combination would lead to excessively large compiled binaries. ATC solves this problem by templating most code only in terms of the floating-point type parameter while using inheritance in some cases to connect components where types are only known at runtime. Although this adds the performance overhead of dynamic dispatch, ATC is specifically designed to only invoke this mechanism in non-critical parts of the code, thus making the performance cost negligible.

\section{Results}
\label{sec:results}

In this section, we will empirically evaluate the performance of ATC in terms of compression rate, speed, memory usage and error control across datasets of different sizes from various application domains. These metrics will be compared to other state-of-the-art numerical data compressors. 

\subsection{Datasets}
\label{sec:datasets}

We selected a diverse collection of datasets across a range of sizes and application domains, as listed in \cref{tab:datasets}. Some of these were already used as benchmarks in other publications \cite{tthresh,sz3-1}. Certain datasets were preprocessed as follows:
\begin{itemize}
\item \textbf{Brain:} The original dataset consists of 7 values per voxel: the 6 unique elements of the corresponding diffusion tensor along with a ``confidence'' value in the interval $[0, 1]$, indicating the accuracy of the measurements for this voxel. We extracted the $6$ actual components and multiplied them by the corresponding confidence value. This effectively removes the distorted values, which represent missing elements, by mapping them to $0$, which makes them easy to compress, while improving smoothness by not applying a hard thresholding scheme.
\item \textbf{Isotropic-PT, Isotropic-V:} These are cut-outs of the ``Isotropic 1024 Fine'' dataset \cite{jhtdb}. Specifically, we extracted the subtensors from the corner with the smallest coordinates in the full tensor. In the case of Isotropic-V, the time mode effectively has size $1$.
\item \textbf{Deforest-8, Deforest-33:} These are part of the ``deforest-globe'' dataset (variant r1i1p1f1) \cite{deforest-globe}. To acquire an appropriate amount of data, we considered only \Verb|.gr| data files with version $20191122$ and annually sampled variables. The filtered data files were then merged into two separate tensors based on their number of levels, i.e. the size of the vertical spatial mode, with $20$ variables using $8$ levels and $19$ variables using $33$ levels. The original data contains extreme but constant values in voxels located over land. Due to a lack of documentation, we assumed that these values represent missing values and should therefore not be stored. Therefore, we subtracted the first time slice from all other time slices. The resulting tensor represents the change in each variable since the start of the simulation, thus setting these missing values to zero and reducing their effect on the compression process. Finally, we normalized each variable by scaling it such that its maximum absolute value equals $1$.
\item \textbf{Hurricane:} To not exceed the available amount of memory on our hardware, we select only the first $20$ time slices. Since the missing values in this dataset were all set to a particular constant, we simply set all of these values to zero and did not apply the difference coding step unlike the previous case. Afterwards, each variable slice was normalized in the aforementioned way.
\end{itemize}

\begin{table}[t]
\small
\centering
\begin{tabular}{|l|l|l|l|l|l|}
\hline
\textbf{Dataset}          & \textbf{Size}          & \textbf{Mode sizes} & \textbf{Data type}                & \textbf{Domain}                              & \textbf{Modes}                                                                     \\ \hline
Foot \cite{foot} & 16.0 MB  & $256 \times 256 \times 256$                 & uint8                    & X-ray scan                          & Space (3)                                                                    \\ \hline
Brain \cite{brain}  & 103 MB & $160 \times 190 \times 148 \times 6$       & float32                  & Diffusion tensor image              & Space (3) $\times$ component                                                    \\ \hline
Moffett-Field \cite{moffett-field} & 224 MB & $512 \times 1024 \times 224$ & int16                    & Hyperspectral image                 & Space (2) $\times$ wavelength                                             \\ \hline
Isotropic-PT \cite{jhtdb}  & 800 MB & $100 \times 128 \times 128 \times 128$ & \multirow{5}{*}{float32} & \multirow{2}{*}{CFD simulation}     & Time $\times$ space (3)                                                    \\ \cline{1-3} \cline{6-6} 
Isotropic-V \cite{jhtdb}  & 1.50 GB & $512 \times 512 \times 512 \times 3$   &                          &                                     & Space (3) $\times$ component                                              \\ \cline{1-3} \cline{5-6} 
Deforest-8 \cite{deforest-globe}   & 3.05 GB & $20 \times 79 \times 8 \times 180 \times 360$   &                          & \multirow{2}{*}{Climate simulation} & \multirow{3}{*}{Variable $\times$ time $\times$ space (3)} \\ \cline{1-3}
Deforest-33 \cite{deforest-globe}  & 12.0 GB & $19 \times 79 \times 33 \times 180 \times 360$  &                          &                                     &                               \\ \cline{1-3} \cline{5-5} 
Hurricane \cite{hurricane-isabel}    & 24.2 GB & $13 \times 20 \times 100 \times 500 \times 500$ &                          & Weather simulation                  &                                      \\ \hline
\end{tabular}

\vspace{5pt}
\caption{The datasets used along with their properties. ``Space (X)'' represents a set of X spatial modes. ``Components'' represents the unique diffusion tensor elements per voxel in the case of Brain and the velocity components per voxel in the case of Isotropic-V.}
\label{tab:datasets}
\end{table}

\subsection{Compressors}
For our experimental comparison, we selected several lossy data compressors based on a recent survey by Duwe et al. \cite[section 1.2.2]{compression-survey}. Specifically, we limited ourselves to compressors with at least support for datasets of order 3 and 4 as well as a publicly available implementation, including a command-line interface, since all experiments are performed using this interface. The resulting compressors are listed in \cref{tab:compressors}. Considering the rate-distortion comparisons performed in \cite{tthresh} and \cite{sz3-1}, this list should contain all relevant compressors. \modified{All compressors, except for the x265 wrapper, were compiled using GCC 8.3.0 \modified[1]{(GNU Compiler Collection)}.}

\begin{table}[]
\small
\centering
\begin{tabular}{|l|l|l|l|l|}
\hline
\textbf{Compressor} & \textbf{Version\modified{/commit date}} & \textbf{\modified{Cores}} & \textbf{Notes}                                                           \\ \hline
ATC        & \modified[4]{1.1.2} & \modified{36} & \modified[1]{Using OpenBLAS 0.2.20 for matrix multiplications.} \\ \hline
TTHRESH    & \modified{20 Feb. 2022} & \modified{36} & \\ \hline
SZ         & \modified{12 Apr. 2022} & \modified{36} & \multirow{3}{0.45\linewidth}{ No support for datasets with 5 or more modes (through the command-line interface). }  \\ \cline{1-3}
ZFP        & \modified{27 Jan. 2022} & \begin{tabular}{@{}l@{}} \modified{36 (compr.)} \\ \modified{1 (decompr.)} \end{tabular} &                                                                 \\ \cline{1-3}
FPZIP      & 1.3.0 & \modified{1} &                                                                 \\ \hline
\modified{TuckerMPI} & \modified{31 Jan. 2022} & \modified{36} & \begin{tabular}{@{}l@{}} \modified{The tensor is sliced into 36 parts along the longest mode.} \\ \modified{Each part is handled by a single-threaded MPI process.} \\ \modified{Since the data is stored in single-precision, we use the} \\ \modified{single-precision driver. Using OpenMPI 2.1.2 with shared-} \\ \modified{memory optimization flags \texttt{--mca btl self,sm,tcp}.} \end{tabular} \\ \hline
x265       & 2.6   & \modified{16} & Based on the x265 video codec \cite{x265}, \modified{see details in text}. \\ \hline
\end{tabular}
\caption{The compressors and their corresponding \modified{information} which will be used in the following compression comparison.}
\label{tab:compressors}
\end{table}

The last compressor in this list, x265, consists of a Python script which preprocesses the dataset and passes it onto the x265 video codec for compression using Fast Forward Moving Picture Experts Group (FFmpeg) version 4.1 and compiled with GCC 6.4.0 \cite{ffmpeg}. We used the options \verb|-preset veryslow -tune psnr|, leading to high compression in terms of \modified{relative error but longer runtimes. Furthermore, the error was controlled using the parameter \texttt{crf}} to optimize compression results. Although we are not aware of any literature considering video compressors for compressing general tensor data, we include x265 in our comparison due to its high performance on video data \cite{netflix-comparison,msu-comparison}.

Because this compressor is intended for video formats, we added preprocessing steps to support compression of general tensors of order 3 or higher. First, for each dataset we designated two of its spatial modes as the video's width and height mode and one mode as the video's time mode. This time mode was chosen as the dataset's own time mode or some other very smooth mode. Then, all other remaining modes were then merged with the time mode by transposing and reshaping the tensor. Finally, all values were shifted, scaled and rounded to fit into the domain $\{0, 1, 2, \dots, 255\}$. Note that due to this limited domain, there will always be a significant quantization error, so our x265-based compressor cannot produce low compression errors and will only be relevant in the high-error domain.

\modified{Finally, while we included FPZIP in all of our experiments, we found that its compression performance was almost always worse than ZFP while being much slower. To reduce the number of data points in the figures in this section, we will therefore not show results from this compressor.}

\subsection{Experimental set-up}

All experiments were performed on a computing node with two Xeon Gold 6240 central processing units (\modified[4]{2.6 GHz clock speed with} 18 physical cores and 36 threads per CPU) and 180 GB main memory running CentOS 7.9.2009. \modified{I/O was performed using local SSD's with theoretical reading and writing speeds of 1 GB/s and 0.6 GB/s respectively.} We let each compressor use \modified{the maximal number of supported threads or processes, which is described in \cref{tab:compressors}}.

Wall-clock times and peak memory usage were measured using the Linux command \verb|/usr/bin/time|. Due to the memory limitations of TTHRESH (see \cref{sec:time-memory}), we will not apply this compressor to the largest dataset, Hurricane. \modified{Furthermore, x265 results will not be shown for Deforest-8 and Deforest-33 since the corresponding compression errors are all larger than $10\%$.}

Throughout this section, we will aggregate average gains in terms of compression rates, speed, memory usage and error control when comparing compressors. These averages refer to geometric means in the case of relative compression gains, speed-ups and peak memory reduction, while in the case of error control we \modified{use the arithmetic mean of the absolute values of the logarithms of the relative error deviations, as shown in the following formula:}
\[
\modified[1]{\text{Average relative error deviation} = \exp \left( \frac{\sum_{i=1}^n \left\lvert \ln \frac{\text{ActualError}_i}{\text{TargetError}_i} \right\rvert}{n} \right) }
\]
Because these metrics are not all evaluated at the same error, we will instead compute the average in between interpolated curves. We obtain these curves by applying linear interpolation to data points consisting of the \modified{logarithm of the compression error}, as well as the peak memory usage or the logarithm of the compression factor, execution time \modified{or deviation from the target error}, depending on the metric under consideration. \Cref{fig:rate-distortion,fig:times,fig:memory,fig:error-control} show that by considering these metrics on a logarithmic scale, we obtain relatively smooth curves. Moreover, we will use this interpolation method when plotting relative compression factors.

Note that some compressors do not accept all data types. Therefore, for datasets with integral data types, we cast them to a $32$-bit floating-point format before compression. After decompression, we round the decompressed values back to the nearest value in the domain of the corresponding integer type, which can significantly reduce the compression error in low-error settings. Note that while ATC performs this rounding step for integral data types internally already, we perform it externally in all cases to prevent this simple feature from skewing our experimental results. Moreover, SZ, ZFP and FPZIP do not support compression of order-$5$ tensors (through the command-line interface), so for these datasets we merge the first two modes together and treat them as order-$4$ tensors.

\subsection{Incremental performance gains compared to TTHRESH}

\modified{\Cref{tab:incremental-performance} shows how each modification to the baseline TTHRESH algorithm affects various performance metrics in our experiments. To clarify the impact of these changes, we briefly discuss our most important findings here:}
\begin{itemize}
\item \modified{\textbf{Baseline ATC:} Due to a more efficient implementation, we achieve a compression speed-up of around \modified[4]{$27$\%} while halving memory usage.}
\item \modified{\textbf{Hybrid truncation:} This leads to compression and decompression speed-up factors of roughly $1.5$ and $1.3$, respectively, at the cost of an $8$\% decrease in compression rate. Note that these numbers are averages and can vary a lot in practice, depending on the compressibility of the dataset and the target error.}
\item \modified{\textbf{Parallel quantization and encoding:} This results in compression and decompression speed-up factors of roughly $1.8$ and \modified[4]{$1.6$}, respectively, while slightly reducing error control. Our experiments show that these speed-up factors are generally present, but can become much smaller for well-compressible datasets and high target errors, since in these cases the truncated core shrinks, reducing the time share of the quantization and encoding phase in the full compression process.}
\item \modified{\textbf{Predicting the dequantization correction during quantization:} This small change greatly improves error control, reducing the average error deviation from $27.7$\% to $1.4$\%, with little effect on other performance metrics.}
\item \modified{\textbf{Various minor improvements:} The three last modifications described in \cref{tab:incremental-performance} slightly improve average compression rate, but \modified[4]{in some cases at a minor speed cost}. As such their relevance may depend on the use case. Therefore, while they are enabled by default, we provide user options to disable them.}
\end{itemize}

\subsection{Rate-distortion comparison}

\begin{figure}[t]
\centering
\input{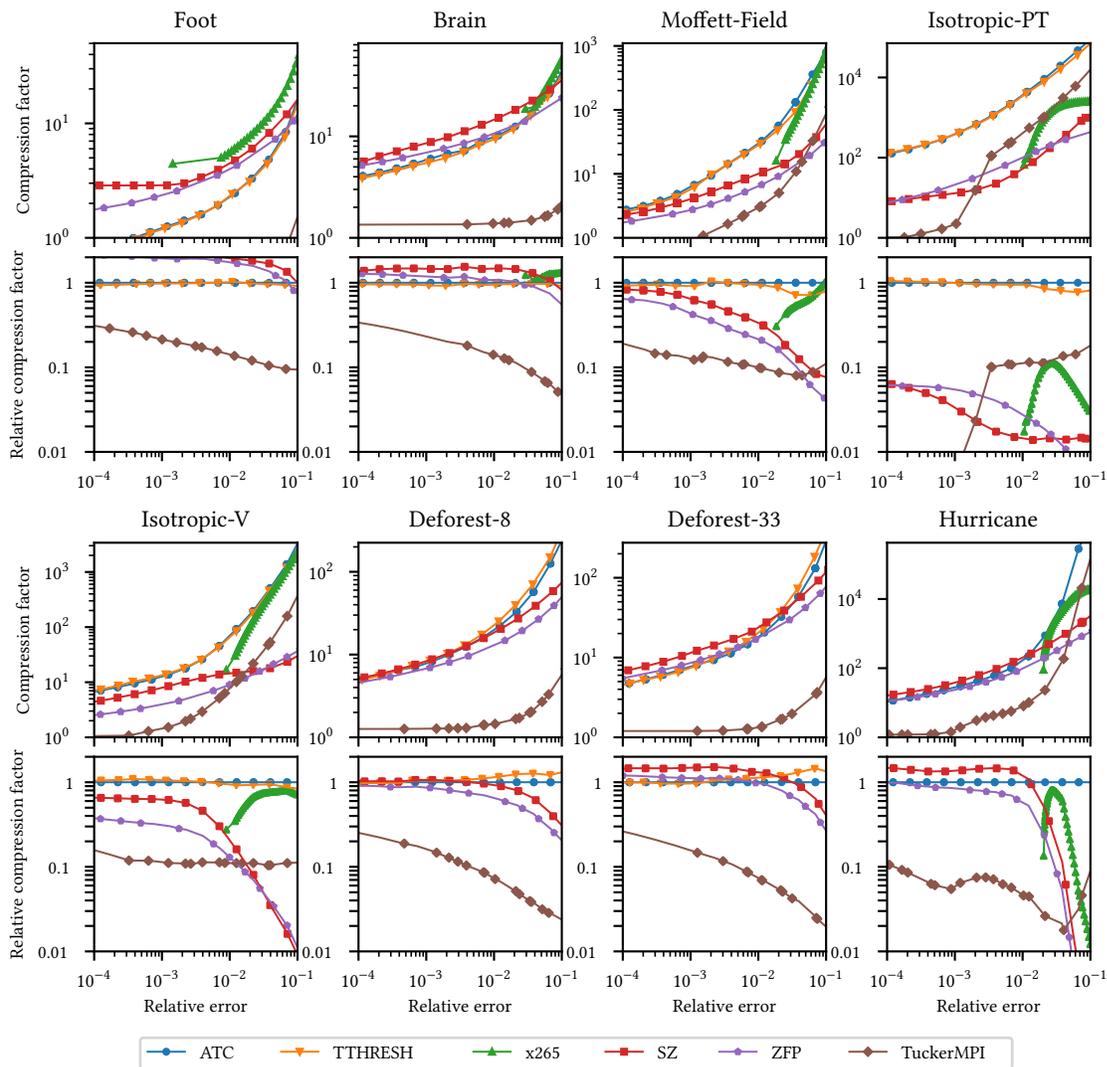}
\vspace{-3em}
\caption{\modified[4]{Rate-distortion curves for all compressors and datasets, both in terms of absolute compression factors and the compression factors relative to ATC.}}
\label{fig:rate-distortion}
\end{figure}

\Cref{fig:rate-distortion} shows the rate-distortion curves for all compressors and datasets, with the highest curves representing the best compression rates. We observed that ATC and TTHRESH perform very well in some cases (Moffett-Field, Isotropic-PT and Isotropic-V), adequate in other cases (Deforest-8, Deforest-33 and Hurricane) and poorly in some other cases (Foot and Brain). In particular, we note that these compressors outperform the rest in the high-compression domain, typically starting from a compression factor of $10$ to $100$. This usually corresponds with high errors, with \modified{ATC/TTHRESH} often achieving \modified{the highest relative compression rates at errors over \modified[1]{1\%}. In fact, while SZ slightly outperforms ATC for many errors, ATC's gain is so high in \modified[1]{other cases} that on average, in our experiments, ATC achieves \modified[4]{98\%} higher compression.}

\begin{figure}[t]
\centering
\input{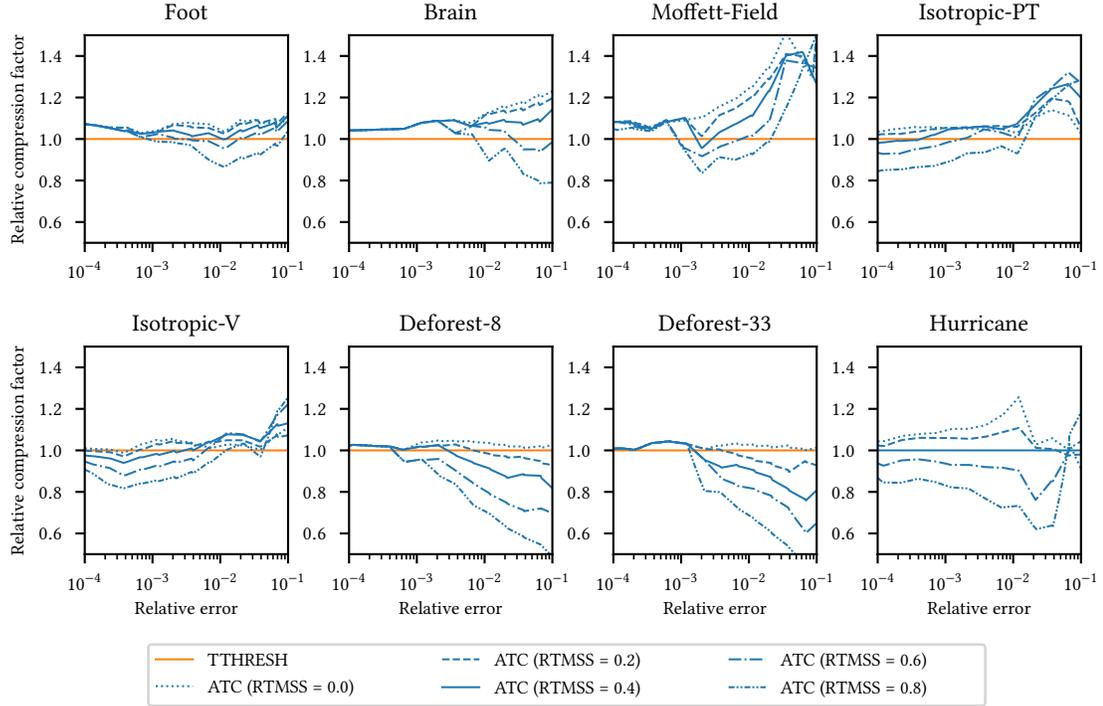}
\vspace{-3em}
\caption{\modified[4]{Compression factors of ATC, in terms of its RTMSS parameter value, relative to TTHRESH. Since we do not have TTHRESH data on Hurricane, we choose ATC with RTMSS $= 0.4$ as a reference instead, which still shows the relationship in between the RTMSS parameter value and the compression factor.}}
\label{fig:rate-distortion-rtmss}
\end{figure}

When comparing ATC and TTHRESH, we can see that both perform similarly most of the time, with ATC only achieving $\modified[4]{1.1\%}$ higher compression on average in our experiments. \Cref{fig:rate-distortion-rtmss} demonstrates that this can be attributed to ATC's default RTMSS parameter value of $0.5$, which controls rank truncation (see \cref{sec:hybrid_truncation}). If we consider the highest compression factor achieved by ATC across the different RTMSS settings for each error value, we obtain an average gain of $\modified[4]{8.0\%}$ compared to TTHRESH, \modified{at the cost of compression and decompression slowdowns of $\modified[4]{24\%}$ and $\modified[4]{21\%}$ respectively \modified[1]{relative to the default RTMSS setting, which is still much faster than TTHRESH}}. In fact, for some datasets this increases to $20\%$ or higher for high errors, which \cref{fig:rate-distortion} showed is the most relevant domain for Tucker-based compressors. Note that this gain is accumulated over the optimal RTMSS setting for each individual error, which is often but not always $0$. Furthermore, we observe that for most errors, decreasing the RTMSS parameter improves compression. In conclusion, we find that users which prioritize compression over speed might want to start with an RTMSS value of $0$ and adjust it if necessary, because this setting almost never leads to compression losses and can lead to significant gains compared to TTHRESH.

\subsection{Time and memory usage}
\label{sec:time-memory}

\begin{figure}[t]
\centering
\input{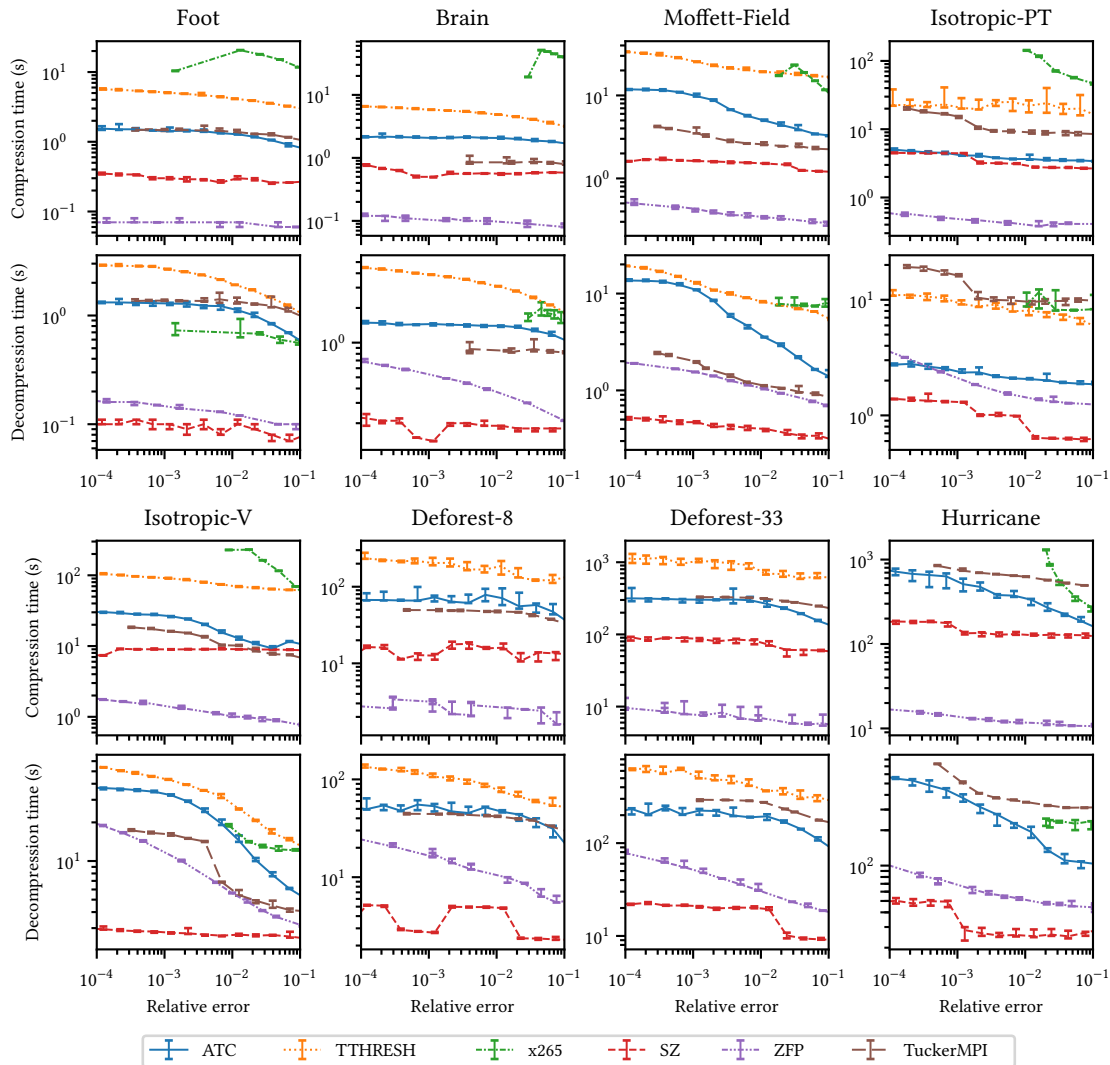}
\vspace{-3em}
\caption{\modified[4]{Median compression and decompression wall-clock times of all compressors, out of $5$ samples per target error. The error bars indicate the range of the measurements among these samples.}}
\label{fig:times}
\end{figure}

\begin{figure}[t]
\centering
\input{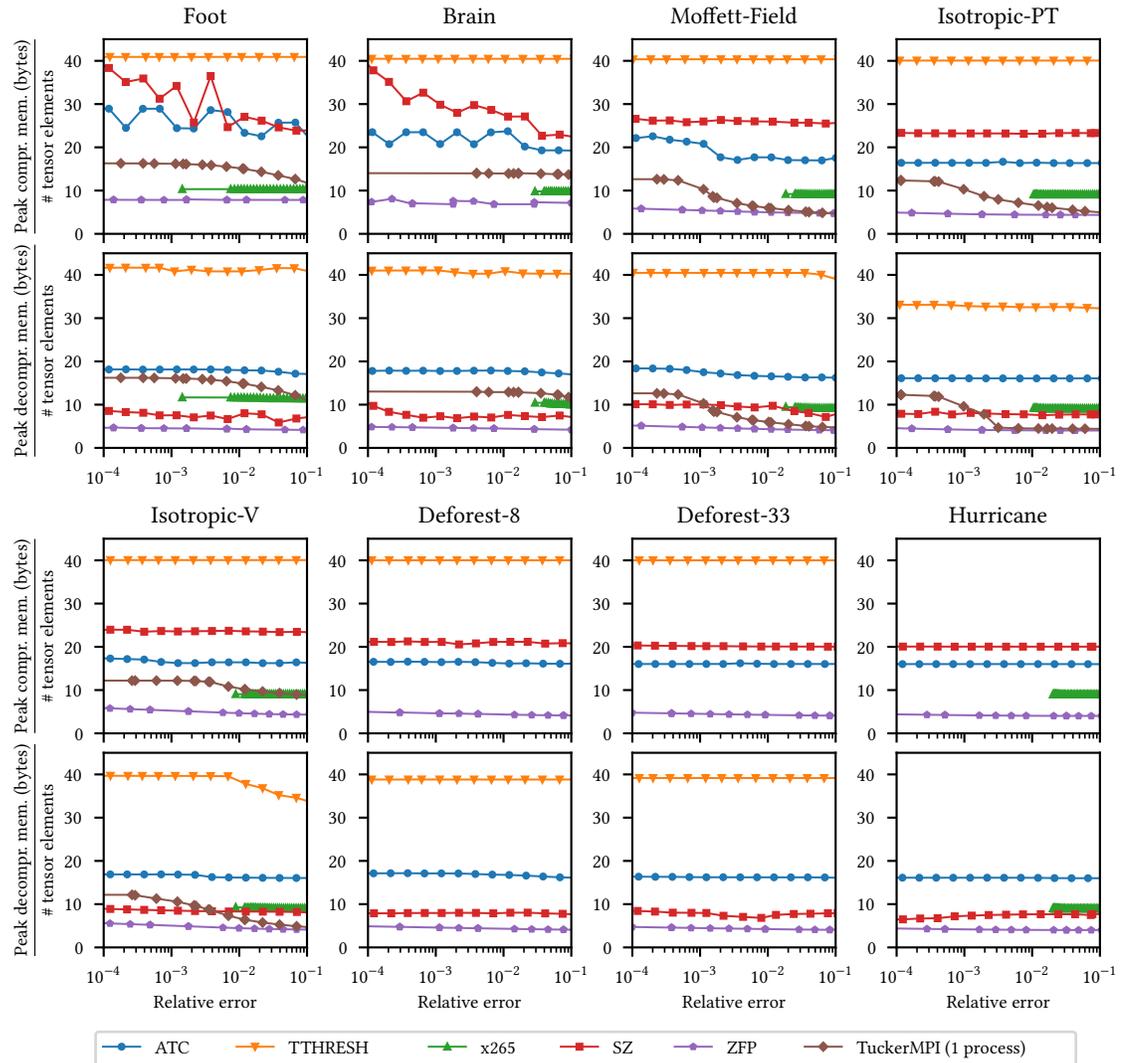}
\vspace{-3em}
\caption{\modified[4]{Peak memory usage of each compressor during compression and decompression. Since this is largely proportional to the number of elements in the original data, we express this metric in terms of the peak number of bytes used per element. To measure TuckerMPI's peak memory usage, we run it using only one process. Because this process needs to read all data alone, it unfortunately cannot process the largest three datasets due to the limitations of MPI I/O.}}
\label{fig:memory}
\end{figure}

\Cref{fig:times} shows the compression and decompression times of each compressor. Note that the Tucker-based compressors generally need more time than the others, \modified{because they apply a global transformation to the data (the Tucker decomposition)}, while most other compressors process the data in blocks in some way, leading to a time complexity proportional to the amount of data. This discrepancy also manifests itself in \cref{fig:memory} in terms of peak memory usage.

However, when comparing ATC to TTHRESH, our compressor took \modified{\modified[4]{$71\%$} and $55\%$} less time on average during compression \modified{and decompression, respectively}. \modified{Yet, we observe that the Tucker decomposition incurs a significant speed cost, making the non-Tucker-based compressors SZ and ZFP much faster in most settings.}

Finally, \cref{fig:memory} shows that ATC is more efficient in terms of memory usage than its Tucker-based counterpart TTHRESH, on average achieving reductions of \modified[4]{$54$\%} and \modified[4]{$56$\%} during compression and decompression respectively. \modified{Yet, note that these averages are skewed by the memory overhead observed in the cases of small datasets like Foot and Brain, while ATC achieves average memory usage of around $18$ bytes per data element for larger datasets, in contrast to TTHRESH's typical $40$ bytes per element.} \modified[2]{Meanwhile, TuckerMPI uses even less memory since we are using its single-precision driver, while ATC stores the tensor in double-precision by default.}

\subsection{Error control}

\begin{figure}[t]
\centering
\input{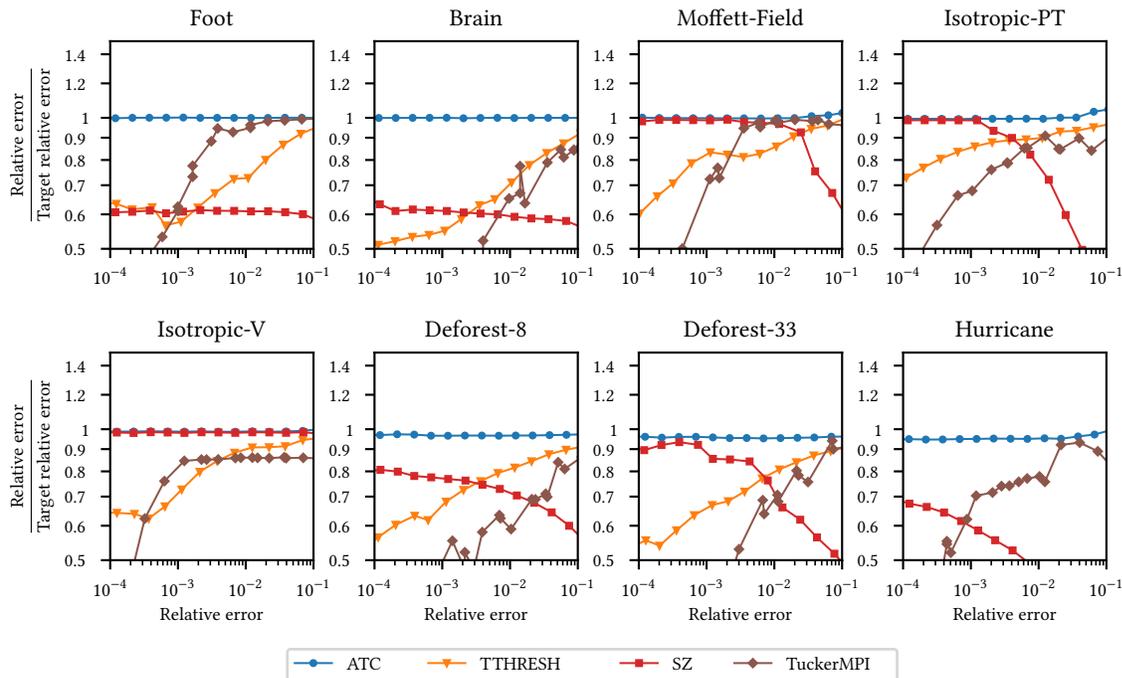}
\vspace{-3em}
\caption{\modified[4]{Error control per compressor, defined as the deviation of the compression error from the target error. Values below 1 indicate that the compressor achieved a lower error than requested. ZFP, FPZIP and our x265 compressor do not support any Euclidean-norm-based target error metric.}}
\label{fig:error-control}
\end{figure}

\modified[1]{Lossy compressors can control the trade-off in between compression rate and error in various ways. In the case of ATC and certain other compressors in our comparison, we say they are \emph{error-bounded} since they attempt to approximate a given target error with the highest possible compression rate.} When using such a compressor, it can be important that the actual compression error does not deviate much from the desired target error. After all, higher errors are undesirable while lower errors lead to an unnecessary loss of compression rate. As a result, we analyzed the degree of error control of different compressors in \cref{fig:error-control}. ATC clearly performs very well, with an average \modified{relative deviation from the target error of $1.4\%$} compared to TTHRESH's \modified{$33.8\%$ and SZ's \modified[4]{$31.7\%$}}.

\section{Conclusions}
\label{sec:conclusions}

We presented ATC, a novel Tucker-based numerical data compressor centered around the ST-HOSVD and bit plane truncation. Several techniques were described to improve \modified{speed, memory usage, error control and compression rate}. Furthermore, certain implementation and usability aspects were discussed.

Our experiments show that ATC on average maintains the compression rates of the state-of-the-art Tucker-based compressor TTHRESH while providing average speed-up factors of \modified[4]{$3.5$} and \modified{$2.2$} during compression and decompression, respectively. Average peak memory usage was also reduced in our experiments by \modified[4]{$54\%$} and \modified{$56\%$} respectively. Moreover, ATC achieves very precise error control, on average only deviating $1.4\%$ from the requested compression error.

Compared to non-Tucker-based compressors, ATC usually outperforms all alternatives in terms of rate-distortion when targeting high \modified[1]{relative} errors\modified{, e.g. above \modified[1]{1\%}.} \modified{Although the state-of-the-art SZ compressor achieves slightly higher compression rates in many settings, ATC drastically outperforms it in some other situations, leading to an average compression gain for ATC of 97\% in our experiments. However, d}ue to the costly Tucker decomposition, ATC uses \modified{significantly} more time and memory\modified{.}

\begin{acks}
\modified{We thank the \modified[5]{five} anonymous reviewers for their extensive questions and remarks on \modified[5]{earlier versions} of this manuscript, which greatly improved this paper.} We are grateful to Rafael Ballester-Ripoll for allowing us to reuse certain parts of the TTHRESH source code in the implementation of ATC. \modified{We kindly thank Zitong Li (TuckerMPI) and Kai Zhao (SZ) for fixing our reported issues, allowing us to run our experiments using their compressors.}

The resources and services used in this work were provided by the VSC (Flemish Supercomputer Center), funded by the Research Foundation---Flanders (FWO) and the Flemish Government. \modified{We thank the VSC support team for extensive help with setting up TuckerMPI on our computing cluster and optimizing the performance of our experiments.}

We \modified{also} thank the sources of all datasets used in this paper: Philips Research and IAPR-TC18 (Foot), Gordon Kindlmann and Andrew Alexander (Brain), the NASA Jet Propulsion Laboratory (Moffett-Field), the Johns Hopkins Turbulence Databases (Isotropic-PT and Isotropic-V), the World Climate Research Programme and the Earth System Grid Federation (Deforest-8 and Deforest-33) as well as the NCAR and U.S. National Science Foundation (Hurricane).

Nick Vannieuwenhoven was partially supported by a Postdoctoral Fellowship of the Research Foundation---Flanders (FWO) with project 12E8119N.
\end{acks}

\bibliographystyle{ACM-Reference-Format}
\bibliography{references}

\end{document}


\title{Minor compression pipeline improvements in ATC}

\author{Wouter Baert}
\email{wouter.baert@kuleuven.be}
\orcid{0000-0002-2846-9199}
\affiliation{
  \institution{KU Leuven}
  \city{Leuven}
  \country{Belgium}
}

\author{Nick Vannieuwenhoven}
\email{nick.vannieuwenhoven@kuleuven.be}
\orcid{0000-0001-5692-4163}
\affiliation{
  \institution{KU Leuven}
  \city{Leuven}
  \country{Belgium}
}

\begin{abstract}
\modified{This document serves as supplementary material to \modified[1]{the} paper, ``\tomsonly{Algorithm xxxx: ATC,}{ATC:} an Advanced Tucker Compression library for multidimensional data''. We describe a \modified[1]{number of additional} techniques implemented in the ATC software package \modified[1]{that lead to more modest performance gains compared to those from the main article.}}
\end{abstract}

\maketitle

\section{Introduction}

\subsection{Outline}

\modified[1]{While our main paper described the most impactful techniques we applied in ATC, we also implemented additional modifications that affect performance to a lesser extent in terms of speed, memory usage, error control and compression rate. First, we describe a scheme based on Householder reflections to improve factor compression (\cref{sec:householder}). Then, we discuss the effects of the way the core is vectorized into the bit matrix for quantization, while proposing a heuristic to determine the optimal mode order the core should be transposed to before vectorization (\cref{sec:core-flattenings}). Afterwards we describe a scheme called split bit plane truncation to improve encoding efficiency on the last bit plane (\cref{sec:split-bit-plane-truncation}), followed by a few improvements to the quantization phase to achieve precise error control (\cref{sec:precise-error-control}). Finally, we implemented an alternative entropy coding method called asymmetric numeral systems \cite{ans} that can be used to slightly increase encoding speed (\cref{sec:entropy-coding}).}

\subsection{Notation}

\modified[1]{In this document, we will use the same notational conventions as in our main paper. We summarize some of the most important ones:
\begin{itemize}
\item The ST-HOSVD \cite{st_hosvd} approximates the original data $\ten{A}$ using a truncated core $\overline{\ten{B}}$ and truncated factors $\overline{U_1}, \dots, \overline{U_d}$.
\item These components are then perturbed during the quantization phase into their quantized versions $\widetilde{\ten{B}}, \widetilde{U_1}, \dots, \widetilde{U_d}$.
\item The decompressed tensor $\widetilde{\ten{A}}$ is reconstructed based on the Tucker decomposition $(\widetilde{U_1}, \dots, \widetilde{U_d}) \tuckerprod \widetilde{\ten{B}}$.
\item The relative error is defined as $\frac{ \| \ten{A} - \widetilde{\ten{A}} \| }{\| \ten{A} \|}$, with $\|\cdot\|$ denoting the Euclidean norm.
\item The sum-of-squared-errors (SSE) is defined as $\| \ten{A} - \widetilde{\ten{A}} \|^2$.
\end{itemize}
}

\section{Householder-based factor compression}
\label{sec:householder}

Although the core takes up most of the compressed output, minor compression improvements can still be achieved by optimizing factor compression. Specifically, we will exploit the fact that each $n \times r$ factor, with $n \geqslant r$, is a matrix with orthonormal columns, i.e. it is subject to $\modified[1]{\binom{r}{2} = r(r - 1)/2}$ orthogonality constraints and $r$ normalization constraints. We can represent these factors succinctly with only $nr - r(r + 1)/2$ coefficients by using \emph{Householder transformations} \cite{matrix-computations}. Namely, we can decompose a truncated factor $\overline{U}$ into $(I - 2 v_1 v_1^T) (I - 2 v_2 v_2^T) \dots (I - 2 v_r v_r^T)$ where each Householder reflector $v_j$ is normalized and $(v_j)_{1:j-1} = 0$. We can conveniently obtain such a factorization by algorithm \ref{alg:householder}, which is an adapted version of the Householder-based QR decomposition algorithm \cite{matrix-computations}. To avoid storing extra signs, the algorithm also adjusts the signs of the factor columns and returns a sign-corrected factor, which can then be used for the corresponding matrix-tensor product in the ST-HOSVD \cite{st_hosvd}.

\begin{algorithm}[t]
\setstretch{0.9}
\SetAlgoLined
\KwData{Factor $\overline{U} \in \mathbb{R}^{n \times r}$}
\KwResult{Householder reflectors in the lower triangle of $V \in \mathbb{R}^{n \times r}$, sign-adjusted factor $\overline{U}' \in \mathbb{R}^{n \times r}$}
 $R = U$\\
 \For{$i = 1, \dots, r$}{
  $v = R_{i:n,i}$\\
  $v_1 = v_1 + \text{sgn}(v_1) \cdot \norm{v}$\\
  $v = v / \norm{v}$\\
  $V_{i:n,i} = v$\\
  $R_{i:n, i:r} = (I - 2 v v^T) R_{i:n, i:r}$\\
  \tcp{If $R_{ii} = -1$, we invert the corresponding column in $\overline{U}'$ such that in the QR decomposition of $\overline{U}' = QR'$, $R'_{ii}$ would be 1, making $R' = I$ so it doesn't need to be stored.}
  $\overline{U}'_{:,i} = \text{sgn}(R_{ii}) \cdot \overline{U}_{:,i}$
 }
 \caption{Converting factor matrices to their underlying Householder reflectors}
 \label{alg:householder}
\end{algorithm}

Each Householder reflector \modified[1]{$v_j$} is normalized so the diagonal elements $v_{jj}$ do not need to be stored, because they can be reconstructed such that $\norm{v_j} = 1$. By construction, $v_{jj}$ is always the biggest component of $v_j$ by absolute value, avoiding any numerical problems. As a result only $nr - r(r + 1)/2$ coefficients need to be stored, which \modified[1]{equals the dimension of the Stiefel manifold of $n \times r$ matrices with orthonormal columns on which $\overline{U}$ lives}.

\modified[1]{Before quantization, the aforementioned coefficients need to be scaled using appropriate weights, because bit plane truncation is based on the principle that an equal absolute error on each quantized coefficient has a roughly equal impact on the compression error. Therefore, we want to assign large weights to the most sensitive scalars in our Householder representation. One approach to this is to scale each reflector $v_j$ with the corresponding slice norm from the quantized core $\widetilde{\sigma}_{j}$, similar to how TTHRESH scales each factor column $U_{:,j}$ with $\widetilde{\sigma}_{j}$ \cite{tthresh}. This way, the factor columns associated with the most important core slices will be reconstructed more precisely. On top of this, we derived a more advanced weighting method that improves error control compared to the aforementioned weights. It is based on a heuristic first-order error analysis tracking of how the error accumulates during the quantization process. With this new weighting, the $i$th element of $v_j$ is scaled by $\sqrt{2\alpha_{ij}}$, where
\[
 \alpha_{ij} = \left( 1 + \frac{n_j + 5\sqrt{n_j} +2}{\bigl(\sqrt{n_j}+1\bigr)^2 \sqrt{n_j}} \right) \widetilde{\sigma}_j^2 + \left( \frac{2}{n_j - 1} + \frac{n_j + 2\sqrt{n_j}+2}{\bigl(\sqrt{n_j}+1\bigr)^3 \sqrt{n_j}} \right) \sum_{k=j+1}^r \widetilde{\sigma}_k^2
\]
and $n_j = n - j + 1$.
}

\section{Alternative core vectorization methods}
\label{sec:core-flattenings}

Usually, most of the compressed output consists of the encoding of the leading bits of the quantized core coefficients. However, due to the run-length and entropy coding applied to this information, the size of this encoding depends on the vectorization method used, i.e. the mapping which flattens the core into a vector. In fact, if all coefficients would be sorted by absolute value in this vectorized core, encoding the leading bits would be trivial and overall compression would at least double in most cases. Although this premise is not realistic, it does show how a different ordering of these coefficients can significantly affect the compression output.

\begin{figure}[t]
    \begin{subfigure}[t]{0.33\textwidth}
        \centering
        \includegraphics[width=0.75\textwidth]{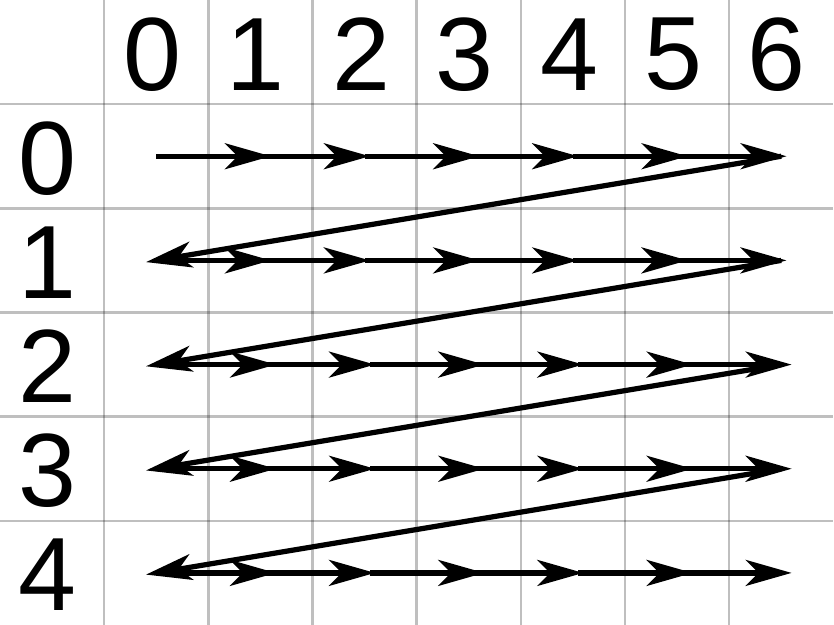}
        \caption{Lexicographic (default)}
    \end{subfigure}
    \begin{subfigure}[t]{0.33\textwidth}
        \centering
        \includegraphics[width=0.75\textwidth]{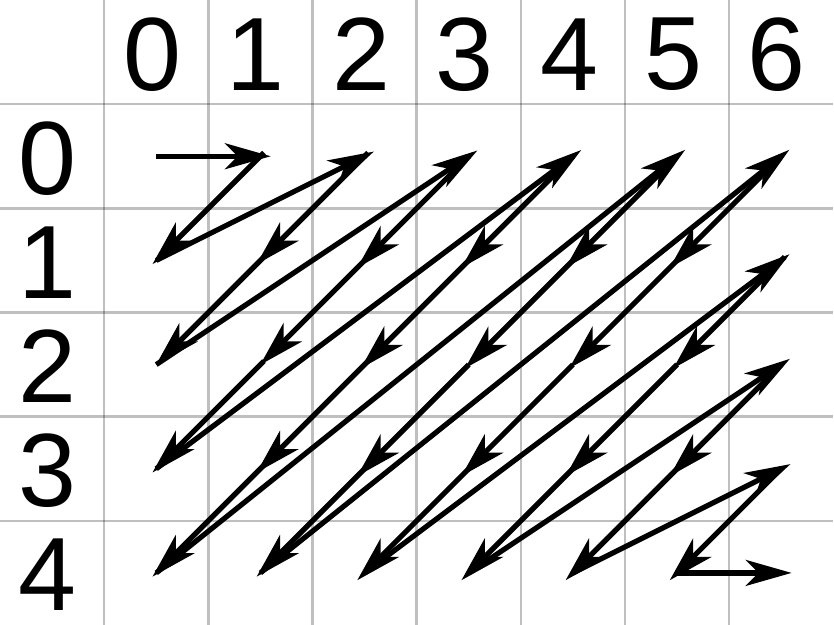}
        \caption{Zigzag}
    \end{subfigure}
    \begin{subfigure}[t]{0.33\textwidth}
        \centering
        \includegraphics[width=0.75\textwidth]{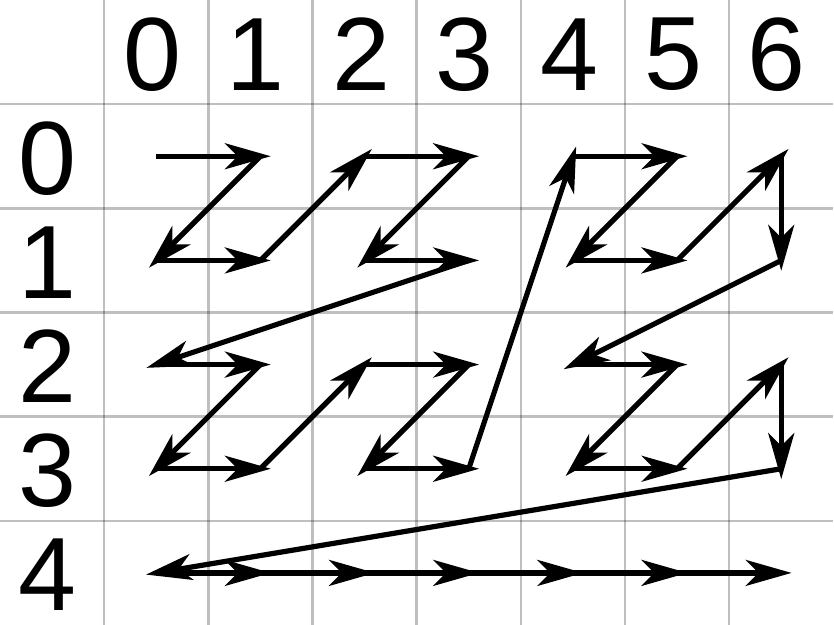}
        \caption{Z-order}
    	\label{fig:flattening-diagrams-z-order}
    \end{subfigure}
    \caption{Different flattenings of a $5 \times 7$ grid.}
    \label{fig:flattening-diagrams}
\end{figure}

In particular, we are interested in ``smooth'' vectorizations which group coefficients of similar absolute value together. This causes the run-lengths to become more extreme, i.e. mostly consisting of very small and very large values, which can be entropy-coded more efficiently than uniformly distributed values. Because of the ``hot-corner'' phenomenon in the ST-HOSVD, the size of the coefficients typically decrease when moving away from the core's origin. Therefore, vectorization methods which preserve locality may be more smooth in the aforementioned sense. To achieve this, we propose the default vectorization strategy, as used in the TTHRESH quantization scheme, as well as two alternatives:

\begin{itemize}
\item \textbf{Lexicographic (default):} All coefficients are ordered in lexicographic order, i.e. with last index changing the fastest. For example, a $2 \times 2 \times 2$ core $\ten{B}$ would be vectorized into $[b_{111}, b_{112}, b_{121}, b_{122}, b_{211}, b_{212}, b_{221}, b_{222}]$. This simply corresponds to a reshape operation and therefore does not require any actual reordering of elements.
\item \textbf{Zigzag:} \modified[1]{Both the JPEG standard and the widely used H.264 video compression standard use a ``zigzag scan'' when processing quantized coefficients \cite{jpeg,h264}.} We generalize the zigzag pattern to higher dimensions as follows: define layer $i$ of the core as the set of positions whose coordinates (0-based) add up to $i$. Then, we first process all coefficients from layer 0, followed by layer 1, and so on. The positions within each layer are processed in lexicographic order. A similar strategy was already proposed for Tucker-based compression in \cite{pre_tthresh}, \modified[1]{but} to the best of our knowledge its performance was never analysed.
\item \textbf{Z-order:} This method was first proposed in 1962 as a file sequencing method \cite{z-order} and is still used e.g. in geospatial databases due to its locality-preserving properties \cite{amazon-aurora}. The Z-order index of a point can be computed simply by interleaving the bits of the binary representations of the point's coordinates into the binary representation of the index. For example, in a 3D grid index 43 corresponds to 101011 in binary, which can be de-interleaved into the binary numbers 10, 01 and 11, leading to the corresponding coordinates (2, 1, 3). \Cref{fig:flattening-diagrams-z-order} shows how points with similar Z-indices are usually close to each other.
\end{itemize}

\begin{figure}[t]
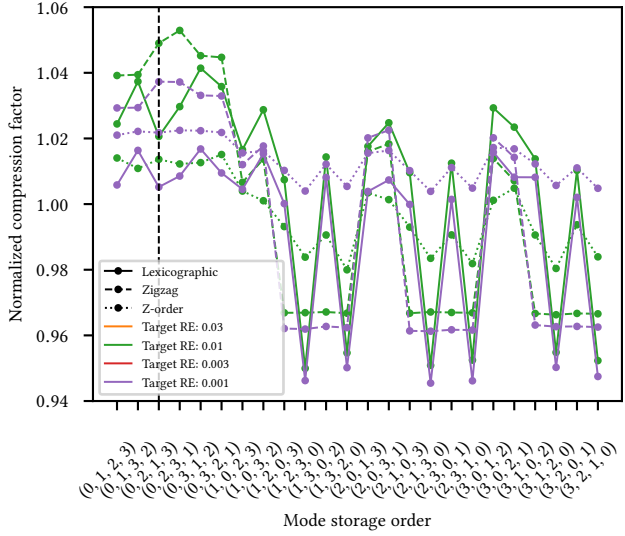

	\begin{minipage}{0.55 \textwidth}
    \begin{subfigure}[t]{\textwidth}
        \input{figures/Isotropic-PT-core-flattenings.pgf}
        \caption{Isotropic-PT. \modified{Core} sizes in terms of target \modified{relative error: \input{data/core_flattenings_core_sizes_Isotropic-PT_filtered}}}
        \label{fig:flattenings-isotropic}
    \end{subfigure}
    \caption{\modified[4]{Compression rate in terms of different vectorization methods, mode storage orders and target errors. To make the comparison of compression rates valid, for each dataset, we only considered target errors for which the actual \modified{errors} for all vectorization methods and storage orders \modified{deviated by no more than 0.1\% from each other}. \modified{Some target errors were also filtered out for the sake of visibility.} Then, for each target error, all compression factors were normalized by dividing them by their collective mean. Therefore, \modified{only} curves which correspond to the same target error (i.e. have the same color) can be compared to each other. \modified{The dashed lines indicate the mode storage order(s) selected by the shortest-mode-first heuristic across all target errors.}}}
    \label{fig:core-flattenings}
	\end{minipage}
	\hfill
	\begin{minipage}{0.42 \textwidth}
	\begin{subfigure}[t]{\textwidth}
		\addtocounter{subfigure}{1} 
        \input{figures/Moffett-Field-core-flattenings.pgf}
        \caption{Moffett-Field. \modified{Core} sizes in terms of target \modified{relative error: \input{data/core_flattenings_core_sizes_Moffett-Field_filtered}}}
        \label{fig:flattenings-moffett-field}
    \end{subfigure}\\
    \begin{subfigure}[t]{\textwidth}
        \input{figures/Foot-core-flattenings.pgf}
        \caption{Foot. \modified{Core} sizes in terms of target \modified{relative error: \input{data/core_flattenings_core_sizes_Foot_filtered}}}
        \label{fig:flattenings-foot}
    \end{subfigure}
	\end{minipage}
\end{figure}

We empirically compared the performance of these vectorization methods across three datasets in \cref{fig:core-flattenings}. For each core, we considered all possible permutations of its modes before vectorizing it. Despite the observations made in \cite{tthresh}, we note that in some settings this mode storage order has a significant impact on the compression rate, with differences of over 10\% in the most extreme cases. This discrepancy can be explained by looking at the corresponding mode sizes: in our situation, short mode sizes usually correspond to highly compressible modes. If these modes are then put at the end of the mode storage order, the size of the coefficients drops very quickly when using lexicographic ordering, leading to a non-smooth flattening. Although the zigzag method processes each layer independently, it is also prone to this effect due to the intra-layer ordering. Conversely, the Z-ordering is \modified{relatively} unaffected by the mode storage order as it never keeps certain mode indices constant over long stretches of the flattening. As a result of these observations, the default mode storage order used in ATC is determined by the order of increasing mode sizes.

When comparing the different flattening strategies, we observe that the default vectorization method and zigzag encoding can lead to noticeably different outcomes, even when following the aforementioned mode storage order heuristic, while Z-ordering provides a stable but suboptimal compression rate regardless of this order. Therefore, ATC supports all three vectorization methods. Lexicographic ordering remains the default option, since it gives the best results along with the zigzag order \modified{on average} in our experiments, while being simpler and therefore slightly faster. Furthermore, all methods require the same amount of memory.

\modified[1]{We} developed and implemented algorithms which can efficiently advance to the next position while staying in-bounds. Specifically, each step takes at most $O(d^2)$ and $O(d \log_2 (\underset{i}{\text{max }} r_i))$ operations for the zigzag order and Z-order respectively. Note that this is an upper bound and most steps take much less time, making the corresponding iterators sufficiently fast. For more information on these algorithms, we refer the reader to \modified{the relevant implementations} in the source file \verb|QuantizedArrayIterator.cpp|.

\section{Split bit plane truncation}
\label{sec:split-bit-plane-truncation}

\begin{figure}[t]
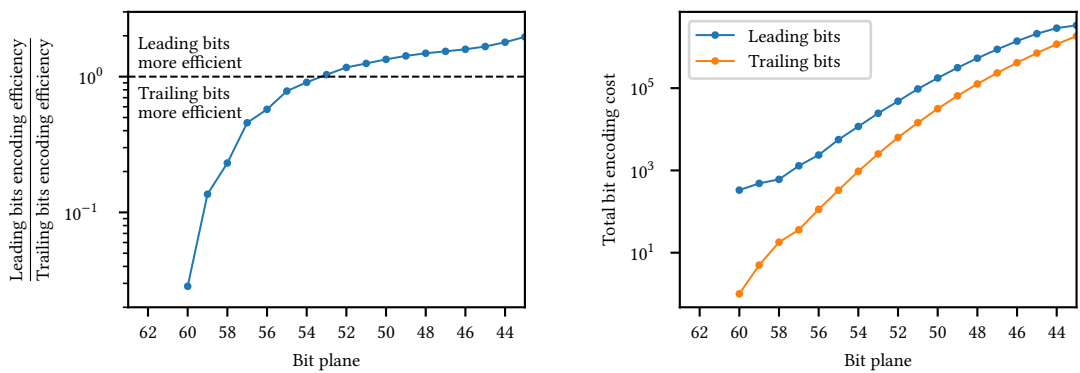

    \centering
    \begin{subfigure}[t]{0.5\textwidth}
        \begin{center}
        	\input{figures/bit-plane-efficiencies.pgf}
        \end{center}
        \caption{\modified{Ratio in between efficiencies for both categories. The encoding efficiency of one bit category is defined as $\frac{\text{SSE reduction}}{\text{Total bit encoding cost}}$.}}
        \label{fig:bit-plane-bit-efficiencies}
    \end{subfigure}
    ~
    \begin{subfigure}[t]{0.5\textwidth}
        \begin{center}
        	\input{figures/bit-plane-bit-costs.pgf}
        \end{center}
        \caption{Total encoding cost for all bits of each category.}
        \label{fig:bit-plane-bit-costs}
    \end{subfigure}
    \caption{\modified[4]{Bit plane encoding statistics obtained by compressing Isotropic-PT at \modified{a target relative error of $10^{-3}$}. Bit plane 63 contains the most significant bits while bit plane 0 contains the least significant ones. The first 3 bit planes are omitted because these results are erratic and irrelevant due to the small amount of encoded information there. \modified{Generally, at least the bit planes down to 57 are encoded.}}}
    \label{fig:bit-plane-statistics}
\end{figure}

The aim of bit plane truncation is to optimize rate-distortion by encoding decreasingly efficient bits from the bit matrix of quantized coefficients. Efficiency in this sense is determined by the error reduction achieved by encoding the bit divided by the encoding cost of the bit. Evidently, encoding bits of higher bit planes takes priority due to their much higher impact on the overall error. Yet, within each bit plane the encoding efficiency may also vary. The encoding cost of a leading bit, i.e. a leading 0 or the first 1 of a quantized coefficient, can be very different from that of a trailing bit\modified[1]{, i.e. a non-leading bit}. Furthermore, the error reduction achieved by encoding such bits is typically far from equal. \modified{In fact, \cref{fig:bit-plane-bit-efficiencies} shows that for many bit planes there is a significant difference in encoding efficiency between both bit categories.} As such, we propose a strategy called \emph{split bit plane truncation}, in which the bit plane truncation breakpoint is split into a separate breakpoint for each bit category. For example, if the leading bits are encoded more efficiently, \modified[1]{\modified{all leading bits} on the last bit plane may be encoded while only some of the trailing bits are, in order to achieve the desired error.} This priority is determined by comparing the efficiencies for both bit categories on the previous bit plane, which is a good predictor of the relative efficiencies on the current bit plane as well\modified{, as demonstrated by the smoothness of the relative efficiency curve in \cref{fig:bit-plane-bit-efficiencies}}.

Since the above optimization only affects the last bit plane, one might think its effect on the compression rate is extremely small. However, note that the encoding cost of each bit plane increases drastically as we progress to less significant bit planes, as demonstrated by \cref{fig:bit-plane-bit-costs}. Therefore, the last bit plane has a relatively large share in the total encoding cost and the overall effect of split bit plane truncation becomes noticeable, although small, in many cases.

\section{Precise error control and recalibration}
\label{sec:precise-error-control}

The basic error tracking method employed by TTHRESH first computes the initial quantization SSE as the squared Euclidean norm of the quantized data (since no bits are encoded yet) and reduces this SSE by the appropriate error reduction each bit plane. While this leads to good error control in most cases, we found that small implementation details could lead to poor error control and erratic encoding breakpoint selection in the case of very low target errors, such as relative errors $\approx 10^{-7}$. We improved the robustness of this method as follows:

\begin{enumerate}
\item \modified[1]{The initial quantization SSE is computed using a naive backward summation.} Because the vectorized core consists of some very large coefficients followed by many small coefficients, this leads to much higher precision than a naive forward summation. Since we already need to loop over all core coefficients to determine the appropriate quantization scaling factor, the SSE computation results in a negligible performance cost. Because of this we chose \modified{backward summation} over the precise but expensive Kahan summation method \cite{kahan-summation} and the fast but less accurate method of computing the core norm based on the ST-HOSVD-truncated singular values.
\item While the initial relative error introduced in the previous step tends to be very small (always under $10^{-10}$ in our experiments but typically much smaller), this can quickly grow as the quantization SSE decreases exponentially in terms of the number of processed bit planes while the initial error remains constant. Therefore, on each bit plane we estimate a reasonable margin for the potential error using results from Robertazzi and Schwarz \cite{summation-error-margin}. If this margin becomes relatively large compared to the current quantization SSE, we explicitly recompute it as $\modified[1]{\sum_i} (\text{quantized\_core[i]} - \modified[4]{\text{current\_encoded\_core[i]}})^2$, \modified[1]{where \modified[4]{current\_encoded\_core} depends on the number of currently encoded bit planes.}
\end{enumerate}

\section{Entropy coding}
\label{sec:entropy-coding}

While ATC's encoding scheme is based on TTHRESH, which uses arithmetic coding, we also support asymmetric numeral systems. As discussed \modified{in \cite{ans}}, the latter's compression rates approach the entropy limit as well, while gaining large to very large speed-ups depending on the variant used.

Since many different run-lengths can occur on the same bit plane, the alphabet of the encoder is usually large which makes variants relying on precomputed look-up tables infeasible. Instead, we implemented the range asymmetric numeral systems variant \cite{rans-tutorial}, which in our experiments processed the run-lengths significantly faster than arithmetic coding, \modified{during both} compression and decompression. However, ATC spends little time on entropy coding compared to the rest of the pipeline, so the overall speed difference is very small. Furthermore, our implementation of range asymmetric numeral systems achieves marginally lower compression rates than arithmetic coding. Because of these reasons ATC supports both entropy coders, with arithmetic coding being the default.

\bibliographystyle{ACM-Reference-Format}
\bibliography{references}